\documentclass{article}
\usepackage[a4paper, total={6in,9in}]{geometry}
\usepackage{mathtools}
\usepackage{authblk}
\usepackage{amsthm}
\usepackage{amsmath}
\usepackage{amsfonts}
\usepackage{amssymb}
\usepackage{hyperref}
\usepackage{natbib}
\usepackage{cleveref}

\theoremstyle{plain}

\newtheorem{definition}{Definition}

\newtheorem{theorem}{Theorem}

\newtheorem{lemma}{Lemma}

\usepackage[utf8]{inputenc}
\PassOptionsToPackage{numbers,compress}{natbib}
\usepackage{algorithm}
\usepackage{algpseudocode}
\usepackage{subcaption}
\usepackage{macros/mylivemacros}

\renewcommand{\algorithmiccomment}[1]{\bgroup\hfill\tiny//~#1\egroup}

\usepackage{tikz}
\usetikzlibrary{external}

\usepackage{pifont}
\usepackage{changepage}

\usetikzlibrary{decorations.pathreplacing}

\newcommand*{\power}[2]{%
  \the\numexpr\poweraux{#1}{#2}\relax
}
\newcommand*{\poweraux}[2]{%
  \ifnum\numexpr#2 = 0 %
    \expandafter\powerend
  \else
    #1 * \expandafter\poweraux
  \fi
  {#1}{#2 - 1}%
}
\newcommand*{\powerend}[2]{1}

\newcommand*{\minimum}[2]{
  \ifnum\numexpr#2 > #1
    #1
  \else
    #2
  \fi
}

\newcommand*{\maximum}[2]{
  \ifnum\numexpr#2 < #1
    #1
  \else
    #2
  \fi
}

\newcommand*{\helper}[3]{
  \ifnum\numexpr#2 < #1
    \maximum{\fpeval{#3 - (#1 - #2)}}{0}
  \else
    \maximum{\fpeval{#3 - (#2 - #1)}}{0}
  \fi
}

\usetikzlibrary{external}
\newif\ifshownotes
\shownotestrue

\newcommand{\blfootnote}[1]{%
  \begingroup
  \renewcommand\thefootnote{}\footnote{#1}%
  \addtocounter{footnote}{-1}%
  \endgroup
}

\newcommand\CoAuthorMark{\footnotemark[\arabic{footnote}]}
\author{Konstantin Donhauser$^*$\protect\CoAuthorMark }
\author{Johan Lokna$^*$}
\author{Amartya Sanyal}
\author{March Boedihardjo}
\author{Robert H\"onig}
\author{Fanny Yang}

\affil{ETH Z\"urich }

\title{Certified private data release for sparse Lipschitz functions}
\date{}
\begin{document}

\maketitle

\begin{abstract}
\blfootnote{* Equal contribution}
As machine learning has become more relevant for everyday applications, a natural requirement is the protection of the privacy of the training data. When the relevant learning questions are unknown in advance, or hyper-parameter tuning plays a central role, one solution is to release a differentially private synthetic data set that leads to similar conclusions as the original training data.
In this work, we introduce an algorithm that enjoys fast rates for the utility loss for sparse Lipschitz queries. Furthermore, we show how to obtain a certificate for the utility loss  for a large class of algorithms.
\end{abstract}

\section{Introduction}
Since sensitive personal information is extensively used in modern data analysis, ensuring the privacy of individual data points has become increasingly critical.
Differential privacy
(DP)~\citep{dwork2006calibrating} attempts to address this issue and 
is used by  
both governmental agencies \citep{Census} and commercial actors \citep{industry_eps}.
Intuitively, a differential private procedure ensures that its output is not affected significantly by individual data points such that it is not possible to determine
whether a particular sample is part of the data set or not. More formally, a probabilistic algorithm $\mathcal{A}$ is said to be $\epsilon$-DP 
if it satisfies the conditions in
Definition~\ref{eq:defdp}. 

\begin{definition}
\label{eq:defdp}
    An algorithm $\gA$ is $\epsilon$-DP with $\epsilon>0$ if for any data sets $D, D'$ differing in a single entry and any measurable subset $S \subset \mathrm{im}(\mathcal A)$ of the image of $\mathcal A$, we have
\begin{equation*}
  \f{\P}{\f{\gA}{D}\in S} \leq \exp\brace{\epsilon} \f{\P}{\f{\gA}{D'}\in S}
\end{equation*}
\end{definition}
A long line of research focuses on preserving differential privacy while extracting specific information from data,
such as executing specific machine learning algorithms~\citep{bassily2014private,chaudhuri2011differentially,feldman2014sample,raskhodnikova2008can}
or answering a number of predetermined queries~\citep{blum2005practical,dagan2022bounded,dwork2004privacy,ghazi21union,hardt2010geometry,steinke2016between,abadi2016deep}
Although showing good performances on downstream tasks, these approaches face the fundamental limitation that  no further queries can be answered after releasing the model without affecting  privacy guarantees. Moreover, the information leakage introduced by hyper-parameter tuning and model selection must be accounted for in the process to avoid a loss in privacy guarantees (see e.g., \cite{papernot2021hyperparameter}).

An approach that mitigates the above shortcomings is to release a synthetic data set in a differentially private manner that is ideally representative of the original data. 
This process is also known as ``data sanitization"~\citep{dwork2009complexity} and has the advantage that any operation performed on the released data set does not introduce further privacy leakage
- a particularly useful property when it is difficult to predict potential future use cases.
Moreover, once a differentially private synthetic data set is generated, any model selection algorithm can be performed on the synthetic data in a regular way.


On a high level, most data sanitization algorithms rely on some sort of discrepancy measure between data sets, which is then approximately minimized by the returned synthetic data set.  A common choice used in SOTA algorithms (see \cite{mst,aim,zhang2017privbayes} and references therein) on downstream tasks 
is to take the Euclidean norm between the histograms of the discretized marginals of the data. This choice for the discrepancy measure, however, does not take the geometry of the underlying space into account. For example, if there is a natural ordering in the domain (e.g. including naturally continuous covariates such as age, income, etc.), DP data generating algorithm could potentially achieve better performance by taking this inherent structure into account.


One natural discrepancy measure that incorporates such a structure is
the Wasserstein distance, studied in recent works \citep{boedihardjo2022private,Yiyun23}. However, the authors show that the required number of samples in the original dataset scales exponentially in the dimension (that is, the minimax optimal rate is of order $n^{-1/d}$). 
An alternative measure is 
the maximum Wasserstein distance of marginal measures over subsets of variables.  Formally,  for an algorithm $\mathcal A$ that takes an empirical distribution $\mu_D$ of a data set $D$  as an input and returns a probability measure $\mathcal A(D)$\footnote{We note that in the literature, the algorithm's output $\mathcal A(D)$ is usually a data set rather than a probability measure. Nevertheless, a data set can always be constructed from a probability measure using standard techniques such as subsampling or discretization (see e.g., \citep{boedihardjo2022private,Yiyun23}). We therefore simplify our notation by disregarding this distinction in our paper. }, we define the following discrepancy measure, also referred to as the   \emph{utility loss} in \cite{boedihardjo2022private}  
\begin{equation}
\label{eq:wassersteinsparse}\utilityloss(\mu_D, \algoD ) :=  \sup_{S \subset [d]; \vert S \vert =s} W_1 \left(P_{\#}^{S}\mu_{D},  P_{\#}^{S}\algoD\right),
\end{equation}
where $W_1$ is the $1$-Wasserstein distance, and $P_{\#}^{S} \mu_{D}$ is the marginal measure of $\mu_D$ on
the $s$-dimensional subspace defined by the coordinates in $S$.  The utility loss measures the transportation cost between the DP measure $\algoD$ and the empirical measure of the private data set $\mu_D$. Moreover, by the   Kantorovich-Rubinstein Duality Theorem \citep{Villani2008OptimalTO}, the utility loss is equivalent to the \emph{maximum mean discrepancy}
\begin{equation}
\label{eq:utilityloss}
    \utilityloss(\mu_D, \mathcal A(D) ) = \sup_{f\in \mathcal F} \left\vert \int f(x) d\mu_D - \int  f(z) d\algoD \right\vert,
\end{equation}
over the function class $\mathcal F$ consisting of all 1-Lispchitz functions (w.r.t.~some metric $\rho$) which are additionally $s$-sparse; that is, $f$ can be expressed as a function that only depends on $s$ dimensions and that is constant over all other dimensions. 
The expression in (2) corresponds to what previous works refer to as the  accuracy or usefulness when the output is a discrete measure.

In this paper, we are the first to answer the following question 
\begin{center}
\emph{Is it possible to generate a private synthetic dataset with a small structured utility loss in Equation~\eqref{eq:wassersteinsparse} from a reasonably sized original datasets}?
\end{center}
A positive answer to this question  would motivate concrete applied future work on approximate implementations. 
In Section~\ref{sec:mainthm} 
we show that there exists indeed an algorithm (Sections~\ref{sec:algo} and \ref{sec:algo_instantioation}) that can achieve a rate of order $n^{-1/s}$ for the utility loss, and thus overcomes the curse of dimensionality. This rate is minimax optimal as a function of $n$, neglecting logarithmic factors.

In order to further enhance the practical utility of our framework, we simultaneously address the question
\begin{center}
\emph{Can we privately output practically meaningful guarantees for the utility loss?}
\end{center}
A tight instance-dependent upper bound (which we call \emph{certificate}) could allow the practitioner to take the maximum utility loss 
into account when deriving insights from the synthetic data.
While Theorem~\ref{cor:rates}  in Section~\ref{sec:mainthm} proves 
an upper bound for the expected utility loss, it only holds for the particular practically infeasible algorithm presented in Section~\ref{sec:algo_instantioation}. Further, evaluating the error bound for a given configuration, i.e. for given $d, n, s$, 
will likely result in a loose, practically meaningless bound.
Instead, we propose an instance-dependent and computable high-probability upper bound that can be privately released alongside the DP measure~$\algoD$. We further show experimentally in Section~\ref{sec:publi_data} that this upper bound is tight.

\subsection{Notation}
\label{subsec:notation}

We refer to data sets of size $n$ with  $D \in T^n$. In the main text of this paper we usually assume that $T = [0,1]^d$ is the hyper cube equipped with the $\ell_{\infty}$-metric and let
 $ T_{k,s} := \{ 1/2k, \cdots , (2k-1)/2k\}^s$ be the centers of a minimal $1/2k$-covering of $[0,1]^s$ of size $N =k^s$. 
We denote with $\measure(T)$ the set of probability measures on $T$ and we let $\mathcal{M}(T)$ denote the set of signed measures on $T$. Moreover, $\mu_D$ is the empirical measure of a data set $D$ and 
 $\mathrm{Lap}(\lambda)$ is the Laplace distribution with zero mean and variance~$2 \lambda^2$. 
 We denote the matrix vector-$1$-norm with  $\|.\|_1$, and use the standard big-O-notation and the symbols $\lesssim_d, \gtrsim_d, \asymp_d$ to hide universal constants only depending on $d$. Finally, we denote with $\triangle^{d-1} \subset \mathbb R^d$ the probability simplex and for any function $g: T \to T'$ mapping $T$ to $T'$, we denote with $g_{\#} :\mathcal{M}(T) \to \mathcal{M}(T')$ the push-forward operator that outputs a signed measure satisfying $g_{\#} \mu(A) = \mu(g^{-1} (A))$ for any $A\subset T'$.








\section{Certified DP data generation}
\label{sec:algo}

In this section, we present a general framework for private data release sketched in Algorithm~\ref{alg:mainalgo}. However, 
 unlike existing approaches, in addition to a DP measure $\mathcal A(D)$, Algorithm~\ref{alg:mainalgo} also returns  a  certificate $\ub_{\proxyutilitytemp}$ for the utility loss - a computable upper bound for the utility loss that holds with probability greater equal $1-\delta$  for some $\delta>0$ and depends on the specific algorithmic choices as well as the particular dataset.

\begin{algorithm}[htbp!]
\caption{Privacy-Preserving Data Generation Framework}\label{alg:mainalgo}
\begin{algorithmic}[1]
\Require Given 
a query operator $\mathbb T$, a  noise generating processes $\prob_{\eta}$ and a  proxy utility loss $\proxyutilitytemp$

\State  project $v \gets \mathbb{T} \mu_D$ \label{lin:projection}
\State construct the $\epsilon$-DP vector $\nudp := v +  \eta$ with $ \eta\sim \noisedist$ 
\label{lin:make_dp}
\State   
$\mudp \leftarrow $ \texttt{minimize}~~$\ub_{\proxyutilitytemp}( \mu, \nudp)$ with respect to $\mu \in \probset$
\label{lin:optimize}
\State \Return DP measure $\mathcal A(D) \gets \mudp$ and certificate $\ub_{\proxyutilitytemp}(\mudp, \nudp)$  for $\utilityloss(\mu_D, \mudp)$
\end{algorithmic}
\end{algorithm}

Following the standard abstract pattern of common  data release frameworks, Algorithm~\ref{alg:mainalgo} consists of three  steps. First, a linear query operator $\queryop: \probset \to \R^{m}$ projects the empirical data distribution of the data set $D$ in the domain $T$, onto a high-dimensional Euclidean space $\R^{m}$. Moreover, let $\image{\mathbb T}$ be the image of $\mathbb T$  and  $\Tinv:  \image{\mathbb T} \to \probset$ be any right-inverse, defined as satisfying  $\mathbb T \Tinv \mathbb T = \mathbb T$. The second step is the standard privatization procedure of adding noise from some distribution $\noisedist$ to the queries. In the third step, we project back from the query space $\R^m$ to the space of probability measures. 
In Algorithm~\ref{alg:mainalgo}, this third step is done by minimizing the upper bound from a DP certificate $\ub_{\proxyutilitytemp} : \probset \times \mathbb R^m \to \mathbb R$ for the utility loss $\utilityloss$. 

In contrast to existing algorithms, instead of solely  releasing the final DP-measure $\mu_{\text{DP}}$, we also output a certificate for the chosen sanitization/generation procedure that we detail in the following.

\paragraph{DP certificate} Before we define our DP certificate, we introduce the concept of a proxy utility loss
\begin{definition}
\label{def:utilityloss} For a right-inverse $\Tinv$, we say that $\proxyutilitytemp: \mathbb R^\Tdim \times \mathbb R^\Tdim \to \mathbb R$ is a proxy utility loss on $\mathbb R^\Tdim$ (\emph{dominating $\utilityloss$}) and for all $v, v' \in \image{\mathbb T} \subseteq \mathbb R^m$, 
\begin{equation}
   \label{eq:UT}
    \utilityloss(\Tinv v, \Tinv v') \leq \proxyutilitytemp(v, v').
\end{equation}
Moreover,  $\proxyutilitytemp$ is jointly translation invariant, that is $ \proxyutilitytemp(v, v') = \utilityloss(v + u, v' + u)$ for any $v, v', u \in \mathbb R^m$, and satisfies the triangle inequality. 
\end{definition}
As a DP certificate $\ub_{\proxyutilitytemp}$, we then propose the following quantity
that can be computed for any probability measure $\mu \in \probset$ and choice of query operator $\queryop$ and proxy utility loss $\proxyutilitytemp$.  
For any $\delta >0$ and any DP-vector $\nudp := \queryop \mu_D + \eta$,
we define \\
\begin{equation}
\label{eq:pluginub}
    \ub_{\proxyutilitytemp}(\mu, \nudp) := \underbrace{\sup_{\tilde \mu \in \probset} \utilityloss(\tilde \mu,\Tinv\mathbb T\tilde \mu)}_{\text{discretization error}}
   +\underbrace{ q_{1-\delta}\left(\proxyutilitytemp(0, \eta)\right)}_{\text{privatization error}}
   + ~\underbrace{\proxyutilitytemp(   \nudp,\mathbb T\mu) + \utilityloss(\Tinv \mathbb T\mu, \mu)}_{\text{projection error}}
\end{equation}
where $q_{1-\delta}(Z)$ denotes the $1-\delta$-quantile of the random variable $Z$ and $\proxyutilitytemp$ is a ''proxy" utility loss, that is, any function $\proxyutilitytemp$ that satisfies the following definition.

In the following lemma, we show that the certificate $\ub_{\proxyutilitytemp}(\mu, \nudp)$ is  a high probability upper bound of the utility loss for any measure $\mu \in \probset$.
\begin{lemma}
\label{lm:upperbound}
For any proxy utility loss  $\proxyutilitytemp$ from Definition~\ref{def:utilityloss} and for $\nudp=\queryop \mu_D+\eta$ with $\eta \sim \noisedist$, we have that with probability $1- \delta$ over $\eta$ (for any $\delta>0$),  it holds that uniformly over all probability measures $\mu \in \probset$, we have
\begin{equation}
\utilityloss(\mu_D, \mu) \leq\ub_{\proxyutilitytemp}(\mu, \nudp).
\label{eq:ub}
\end{equation}
\end{lemma}

\paragraph{Proof}  
Using the triangle inequality and Equation~\eqref{eq:UT}, we can upper bound the utility loss \eqref{eq:utilityloss} for any $\mu \in \probset$ and $\mu_D$ by
\begin{align}
   \utilityloss(\mu_D, \mu) 
   &\leq \utilityloss(\mu_D,\Tinv \mathbb T\mu_D)+\utilityloss(\Tinv \mathbb T\mu_D, \Tinv \mathbb T\mu) + \utilityloss(\Tinv \mathbb T\mu, \mu)
   \nonumber \\
      &\leq \utilityloss(\mu_D,\Tinv \mathbb T\mu_D)+\proxyutilitytemp( \mathbb T\mu_D,  \mathbb T\mu) + \utilityloss(\Tinv \mathbb T\mu, \mu)
   \nonumber \\
   &\leq \utilityloss(\mu_D,\Tinv \mathbb T\mu_D)+\proxyutilitytemp( \mathbb T\mu_D,  \nudp )+\proxyutilitytemp(   \nudp,  \mathbb T\mu) + \utilityloss(\Tinv \mathbb T\mu, \mu)\nonumber\\
   &= \utilityloss(\mu_D,\Tinv \mathbb T\mu_D) + 
\proxyutilitytemp(0,   \eta ) + 
\proxyutilitytemp(    \nudp, \mathbb T\mu) + \utilityloss(\Tinv \mathbb T\mu, \mu)
\label{eq:general_ub123}
\end{align}
where in the last equality we used that $\nudp = \queryop \mu_D + \eta$ and the joint translation invariance of $\proxyutilitytemp$.

The first term can be described as the discretization error associated with $\Tinv \mathbb T$ and can be bounded by the supremum over all probability measures. 
Since we know the noise distribution $\noisedist$, the second term can be upper bounded with high probability by its $(1-\delta)$-quantile.
Finally, the third term only depends on $\mu$ and $\nudp$, which is DP by construction, and the fourth term only on $\mu$. 
As a result, we obtain the desired DP upper bound from Equation~\eqref{eq:pluginub}.\qed


\section{Instantiation of the algorithm for the sparse Wasserstein loss}
\label{sec:algo_instantioation}

In this section we now present an instantiation of Algorithm~\ref{alg:mainalgo} for the utility loss in Equation~\eqref{eq:wassersteinsparse}. The presented algorithm has an exponential run-time complexity in $n$ (see discussion in Section~\ref{sec:mainthm}) and we leave practically useful approximate algorithms as a future work. 
For simplicity of exposition, throughout this section we only consider the case where the underlying space $T=[0,1]^d$ is the hypercube equipped with the $\ell_{\infty}$-metric and refer to reader to Appendix~\ref{sec:apxmainthm} for results on general metric spaces. We first 
define the specific choice of the query operator $\mathbb T$, the noise generating processes $\prob_{\eta}$ and the proxy utility loss $\proxyutilitytemp$. Finally, we summarize Algorithm~\ref{alg:mainalgo}.



\paragraph{Query operator $\mathbb T$}
\label{subsec:t}
Similar to previous works (see e.g.,  \citep{aim,mst,zhang2017privbayes}), we choose $\mathbb T \mu_D$ to be the vector representing all discretized $s$-marginals.  More precisely, the output of the  query operator $\mathbb T:  \probset \to  \mathbb R^{{d \choose s} k^s}$ consists of all ${d \choose s} $ blocks $\mathbb T \mu_D =: v = [v^{S_1}, \cdots, v^{S_{K}}]^T$ with $K = {d \choose s}$ of size $v ^{S_j}\in \mathbb R^{k^s}$ and $k \in \mathbb N_+$ is some discretization parameter. Moreover, $S_j \subset [d]$ and $\vert S_j\vert = s$ are all subsets of size $s$.

Every block is constructed by first projecting the measure $\mu$ on its marginals $\mu^{S_j} = P_{\#}^{S_j} \mu$, which we then  discretize by a finite measure on $T_{k,s}$   forming a $1/2k$-covering of $P^{S_j} T$ of size $k^s$.
Finally,  we choose any (arbitrary) right-inverse $\Tinv : \image{\mathbb T} \to \mathcal{M}_{\mathbb P}(T_{k,d}) \subset \probset$  such that it returns a finite measure on $T_{k,d}$. 




\paragraph{Noise vector $\eta$}
\label{subsec:eta}
As in \citep{boedihardjo2022private,Yiyun23}, we use the (matrix transformed) Laplace mechanism \citep{dwork2006calibrating,xiao2010differential}, which generates a DP ``copy'' $\nudp$ (Step 2 in Algorithm~\ref{alg:mainalgo}) of the vector $v := \mathbb T\mu_D$ (Step 1 in Algorithm~\ref{alg:mainalgo}) by adding matrix transformed i.i.d~Laplace noise\footnote{
Since all entries of $\mathbb T\mu_D$ are multiples of $\frac{1}{n}$, an alternative choice would be to use the discrete Laplace mechanism \citep{Yiyun23, discretelaplace}, which yields the same theoretic guarantees in Section~\ref{sec:mainthm} when straight forwardly modifying the proofs in Appendix \ref{sec:apxmainthm}.} 
\begin{equation}
\label{eq:noiseeq}
    \nudp := v + \eta := v +  \left[\Phi \tilde \eta\right]_{1:\Tdim}, 
\end{equation}
where $\tilde \eta$ is an i.i.d.~Laplace random vector with variance  as in Lemma~\ref{lm:privacy_phi} and $\Phi \in \mathbb R^{\Tdim_{\Phi} \times \Tdim_{\Phi}}$ is some
invertible matrix of dimension $\Tdim_{\Phi} \geq \Tdim$. A standard quantity when comparing two datasets is the sensitivity of $\mathbb T$ 
\begin{equation}
      \Delta_{\mathbb{T}} := \sup_{D, D'} \| \mathbb{T}\mu_D - \mathbb{T}\mu_{D'} \|_1,
      \label{eq:sensitivity}
\end{equation}
where we take the supremum over all datasets $D, D' \subset   T^{n}$ of size $n$ which differ in at most one point.
Using the previous definitions, the following privacy guarantee holds:

\begin{lemma}(Corollary of Theorem 3.6 in \citep{gauss_mechanism}) \label{lm:privacy_phi}   The vector $\nudp$ is $\epsilon$-DP for $\eta =  \left[\Phi \tilde \eta\right]_{1:\Tdim}$ and
\begin{equation}
    \tilde \eta \sim \left(\mathrm{Lap}\left(
\frac{\|\Phi^{-1}\|_1 \Delta_{\mathbb
T}}{\epsilon}\right)\right)^{\Tdim_{\Phi}}.
\end{equation}
\end{lemma}

By multiplying  the noise in Equation~\eqref{eq:noiseeq} with the matrix $\Phi$, we obtain a ``correlated'' noise. While such a  noise has been previously used in the literature (see e.g., \citep{privelet}), a key insight in the paper \citep{boedihardjo2022private} is to use a Haar-matrix transformed Laplacian noise to obtain tight guarantees for the utility loss of $1$-Lipschitz continuous functions (i.e., when $s=d$).  Based on this idea, we now describe our choice of $\Phi$ used in Step 2 of Algorithm~\ref{alg:mainalgo}.  
Recall that we denote with $v^{S_j} \in \mathbb R^{k^s}$ the $j$-th block of the vector $v$. For every $j \in[{d \choose s}]$ we define $\nudp^{S_j} := v^{S_j} + \left[\Phi^{S_j} \tilde \eta^{S_j}\right]_{1:k^s}$
where 
$\Phi^{S_j}$  are  the scaled versions of the transposed Haar-matrix  from Lemma~\ref{lm:haar_properties} in Appendix~\ref{apx:haar} such that $\|(\Phi^{S_j})^{-1}\|_1=1$.
Furthermore, $\tilde \eta^{S_j}$ are i.i.d.~Laplacian random vectors with variance as in Lemma~\ref{lm:privacy_phi} and $\Delta_{\mathbb T} ={ d \choose s} \frac{2}{n}$.

\paragraph{Proxy utility loss $\proxyutilitytemp$}
\label{subsec:proxy}

We now describe a choice for a proxy utility loss $\proxyutilitytemp = \utilitylossT$ such that the certificate $\ub_{\utilitylossT}$ can be effectively minimized using linear programming (see Section~\ref{sec:experiments} for a computationally efficient approximation).
 Inspired by  \citep{boedihardjo2022private} which studies the case where $s=d$, we choose 
\begin{equation}
   \utilitylossT(v, u) :=  \max_{S \subset [d], \vert S \vert =s  } \frac{1}{k} \sum_{l=1}^{k^s} \left\vert \sum_{i=1}^l v^{S}_i -  u^{S}_i\right\vert.
         \label{eq:defofH}
\end{equation}
where $v^{S}_i$ is the $i$-th element of the block $v^S \in \mathbb R^{k^s}$. We assume that the elements of $\nudp$ (and thus also $\mathbb T \mu_D$) are ordered as follows:
 first note that any block vector $\nudp^{S_j} \in \mathbb R^{k^s}$ of  $\nudp$  has a one-to-one mapping to a  discrete signed measure
$\omega \in \mathcal{M}_{\mathbb P}(T_{k,s})$ on the $s$-dimensional discrete hyper cube $T_{k,s}$, 
defined by $\omega^{S_j}_{\nudp} := \underset{z_i \in T_{k,s}}{\sum}  (\nudp^{S_j})_i\delta[z_i]$, where $\delta$ is the Dirac-delta point measure. Using this definition, we order the indices of $v^S_i$ such that the corresponding centers  $z_i$ form a Hamiltonian path, or more formally, such that for all $i \leq k^s-1$,  $\| z_i - z_{i+1} \|_{\infty} = 1/2k$.


We refer to Appendix~\ref{apx:generalproofk1} and \ref{apx:generalproofkall} for a proof that $\utilitylossT$ indeed satisfies Definition~\ref{def:utilityloss}.  On a high level, the Hamiltonian path allows us to reduce the problem of constructing a proxy utility loss function over vectors $v^{S}$ representing discrete measures in a $s$-dimensional space to one of construction a proxy utility loss function over discrete measures on an interval of $\mathbb R$.

\paragraph{Minimization of $\ub_{\proxyutilitytemp}(\mudp, \nudp)$ in Step 3}
\label{subsec:algo}
We finally describe how we can minimize $\ub_{\proxyutilitytemp}(\mudp, \nudp)$ in Step 3 in Algorithm~\ref{alg:mainalgo}. 
Note that because the query operator $\mathbb T$ discretizes every marginal measure using a $1/2k$-covering,  the maximum discretization error in Equation~\eqref{eq:pluginub} equals
\begin{equation}
  \sup_{\tilde \mu \in \probset} \utilityloss(\tilde \mu,\Tinv\mathbb T\tilde \mu) = 1/2k. 
\end{equation}
Moreover, the term $\utilityloss(\Tinv \mathbb T\mu, \mu)$ is zero whenever $\mu \in \probsetk = \Tinv \mathbb T \mathcal M(T)$. Thus, minimizing the upper bound from the certificate $\ub_{\utilitylossT}$ in Step 3 in Algorithm~\ref{alg:mainalgo} simplifies to 
\begin{equation}
\label{eq:step3}
\arg\min_{\mu \in \probsetk} \utilitylossT(\nudp, \mathbb T \mu).
\end{equation}
 Finally, we can output the certificate in Step 4 in Algorithm~\ref{alg:mainalgo} after computing the $1-\delta$ quantile of $\utilitylossT(0,\eta)$, which we can efficiently approximate using Monte Carlo samples.

\section{Statistical rates for the utility loss}
\label{sec:mainthm}
In this section, we present rates for the expected utility loss of the instantiation of Algorithm~\ref{alg:mainalgo} as described in Section~\ref{sec:algo_instantioation}. 
To the best of our knowledge, we are the first to study the utility loss in Equation~\eqref{eq:wassersteinsparse}. 
Theorem~\ref{cor:rates} shows
that the rate of the utility loss only depends on $s$ in the exponent but not on $d$.  Thus, we see that by only considering $s$-sparse marginals in Equation~\eqref{eq:wassersteinsparse} we can effectively  overcome the curse of dimensionality. 

\begin{theorem}
\label{cor:rates}
Let $\mathcal F$ be the set of $s$-sparse $1$-Lipschitz  functions on $T=[0,1]^d$ with respect to the $\ell_{\infty}$-metric. Then, for any $n\epsilon \geq d$ and any $s\leq d$, Algorithm~\ref{alg:mainalgo}  is $\epsilon$-DP and for ${k \asymp d^{-1} \left( \frac{ \log(\epsilon n)^{2}}{n \epsilon}\right)^{-1/s}}$  has an expected utility loss \eqref{eq:utilityloss} at most
\begin{equation}
\label{eq:ratessparse}
    \mathbb E ~\utilityloss (\mu_D, \algoD) \lesssim_s  ~d \left( \frac{ \log(\epsilon n)^{2}}{n \epsilon}\right)^{1/s}.
\end{equation}
\end{theorem}
We refer to Appendix~\ref{apx:main} for the proof of Theorem~\ref{cor:rates} which is a consequence of the general statement Theorem~\ref{thm:mainthm} that applies to general metric spaces. 
In fact, the rate in Theorem~\ref{cor:rates} is optimal in $n$ up to logarithmic factors. Indeed, as a corollary of the results in Section 8 in \citep{boedihardjo2022private} we derive the following information theoretic lower bound on the expected utility loss:
\begin{equation}
\label{eq:lowerbound}
    \inf_{\algoD: \epsilon\text{-DP}} \sup_{D\in T^n}~ \mathbb E ~\utilityloss(\mu_D, \mathcal A(D)) \gtrsim   \left( \frac{ \floor{d/s}}{n \epsilon}\right)^{1/s}
\end{equation}

\paragraph{Open problem: tightness in $d$}
While this lower bound matches the exponential decay rate in $n$ of the upper bound in Theorem~\ref{cor:rates}, the dependency on $d$ is not the same. 
Nonetheless, tightening this gap poses a challenging problem and we believe it will require novel creative ideas. The main difficulty arises from the interdependence of the ${ d \choose s}$ marginal measures, which makes it challenging to enhance either of the two bounds without carefully considering this dependency.
We motivate future work to solve the open problem of 
finding the right dependency on $d$ in the lower bound in Equation~\ref{eq:lowerbound}  supported by a matching upper bound.



\paragraph{Proof sketch}
The proof of Theorem~\ref{cor:rates} builds on the ideas developed in \citep{boedihardjo2022private} for the case where $s=d$. While the result is a relatively straight forward extension, the main technical contributions are two-folds:  we simplify the proofs in the mentioned paper and present them in the context of Section~\ref{sec:algo} (see Appendix~\ref{apx:generalproofk1}), 
 which then allows us to extend the results in \cite{boedihardjo2022private} to the case where $s <d$ (see Section~\ref{apx:generalproofkall} and \ref{apx:main}).  

The first part of the proof is devoted to 
showing that $\utilitylossT$ indeed satisfies Definition~\ref{def:utilityloss}. Using Equation~\eqref{eq:general_ub123} (with $\proxyutilitytemp = \utilitylossT$), we then can  upper bound the utility loss in expectation by 
\begin{align}
   &\mathbb E~\utilityloss(\mu_D, \mathcal A(D)) 
   \nonumber\\
   &\leq
\utilityloss(\mu_D,\Tinv\mathbb T\mu_D)+ \mathbb E~\utilitylossT(\mathbb T\mu_D,\nudp)+ \mathbb E~\utilitylossT(\nudp,\mathbb T\algoD) + \utilityloss(\Tinv \mathbb T\algoD, \algoD) \nonumber 
\\
&\leq 1/2k + 2~ \mathbb E~\utilitylossT(\mathbb T \mu_D, \nudp) 
  \label{eq:utilityUBgeneral}
\end{align}
where we used the
fact that $\algoD$ is a
solution of Equation~\eqref{eq:step3} and thus ${\utilityloss(\Tinv \mathbb T\algoD, \algoD) =0}$ and $  \utilitylossT(\nudp,\mathbb T\algoD) \leq \utilitylossT(\mathbb T\mu_D,\nudp)$, and that the discretization error is upper bounded by $1/2k$. Thus, we obtain an upper bound for the expected utility loss when bounding the term $\mathbb E ~\utilitylossT(\mathbb T \mu_D, \nudp ) =  \mathbb E~\utilitylossT(0, \left[\Phi \tilde \eta\right]_{1:\Tdim})$, which only depends on the random vector $\tilde \eta$ and $\Phi$, but not on the measure $\mu_D$. In this step, we crucially rely on the choice of $\Phi$ in Section~\ref{subsec:algo}. Finally, we obtain the bound in Theorem~\ref{cor:rates} by optimizing over the discretization parameter $k$.

\subsection{Further discussion}

\paragraph{Run-time complexity}
We now discuss the run-time complexity of Algorithm~\ref{alg:mainalgo}. First note that both Step 1 and 2 in Algorithm~\ref{alg:mainalgo} have a run-time complexity of $O(d^s n + m) = O(d^s n + d^s k^s)$ (and thus polynomial in $d$).
However, solving the minimization problem in Step 3 in Algorithm~\ref{alg:mainalgo} requires to run a linear program over the $\vert T_k \vert = k^d$ free variables, which has a run-time complexity of $O(\text{poly}(m + k^d)) =O(\text{poly}(d^s k^s + k^d))  $.

In particular, when pluging-in the optimal choice for the discretization parameter $k$ from Theorem~\ref{cor:rates}, we obtain a run-time complexity of order $O(d^s n + \text{poly}(n^{d/s}))$. For small constant choices of $s$ we therefore obtain an exponential run-time complexity in $d$, and computational
hardness results \citep{dwork2009complexity,ullman2011pcps} for the special case of estimating $2$-way marginals in fact suggest
that the exponential dependency in $d$ cannot be avoided.  Nevertheless, we can still hope for practically meaningful approximate algorithms with fast (polynomial) run-time complexity (see Section~\ref{sec:publi_data}) and motivate future work on this topic.


\paragraph{Other types of sparsity} In Theorem~\ref{thm:mainthm} we obtain fast rates without a dependency on $d$ in the exponent by restricting $\mathcal F$ to $s$-sparse Lipschitz functions. As we show in Appendix~\ref{apx:measure}, we can also obtain similar results when $\mathcal F$ is the set of all $1$-Lipschitz functions but the data itself lives on a (unknown) $s$-dimensional space. Importantly, the algorithm can adapt to the degree of the sparsity and does not need to have access to the effective dimension $s$ of the data. This opens up the pathway for adaptive DP data generating algorithms, which we leave as future work.

\paragraph{Comparison with \citep{boedihardjo2022private,Yiyun23}} Previous works considered the special case where the utility loss is the Wasserstein distance, i.e. the loss from Equation~\eqref{eq:wassersteinsparse} for $s=d$. 
In this case, the authors show that the optimal rate for the utility loss is of order $n^{-1/d}$ (see Equation~\eqref{eq:ubfull} in Section~\ref{sec:relatedwork}). 
Theorem~\ref{cor:rates} shows that by restricting to $s$-way marginals, we can address the curse of dimensionality in the rates for the utility loss, resulting in only a linear dependency in $d$ instead of an exponential.





\section{A tighter certificate and numerical evaluation using public data}
\label{sec:publi_data}
In this section we  present numerical simulations illustrating the utility loss $\utilityloss$ and the certificate $\ub_{\utilitylossT}$ from Section~\ref{sec:algo_instantioation} for  Algorithm~\ref{alg:mainalgo}. To avoid the exponential run-time complexity, we first introduce in Section~\ref{subsec:pudata} a computationally efficient approximation of Algorithm~\ref{alg:mainalgo} using public data.
We then present in Section~\ref{sec:experiments} the numerical simulations on real-world data sets and further present 
in Section~\ref{subsec:tightcertificate}  a tighter choice for the proxy utility loss $\proxyutility$ which yields a tighter certificate. The numerical analysis presented in this paper serves as a proof of concept and motivates future research on efficient approximate algorithms.

\subsection{Computationally efficient approximation of Step 3 in  Algorithm~\ref{alg:mainalgo}}
\label{subsec:pudata}
To avoid the exponential run-time complexity of Step~3 in Algorithm~\ref{alg:mainalgo} we  restrict the search space for $\mu_\text{DP}$ to a 
set $\approxmeas(T_k)$ of discrete measures that are supported on a given public data set $\pubdata = \{ z_i\}_{i=1}^{\pubm }  \subset T_k$.  We can then approximate Step 3 in Algorithm~\ref{alg:mainalgo} (using Equation~\eqref{eq:step3}) by
\begin{equation}
   \mu_{\text{approx}} \in \arg\min_{\mu \in \approxmeas(T_k)}~\utilitylossT(\nudp,  \mathbb T \mu)~~ \: \text{with} \:~~
    \mathcal \approxmeas(T_k) = \left\{ \sum_{i=1}^{\pubm} \alpha_i \delta[ z_i]~ \vert~ \alpha \in \Delta^{\pubm - 1}\right\}.
    \label{eq:pubdata}
\end{equation}

The approach of using a public data set  has been previously proposed in the literature \citep{boedihardjo2021privacy,liu2021leveraging} to improve the computational efficiency. In fact, the  optimization problem in Equation~\eqref{eq:pubdata} 
is still a linear program and can be solved with run-time complexity 
$O(\text{poly}(d^s \pubm + m)) = O(\text{poly}(d^s)~\text{poly}(\pubm + k^s))$. Thus, we obtain a polynomial dependency on $d$ given that $\pubm$ grows at most polynomially in $d$. 

\begin{figure}[t!]

    \centering
    \begin{subfigure}{\textwidth}
        \centering
\includegraphics[width=\textwidth]{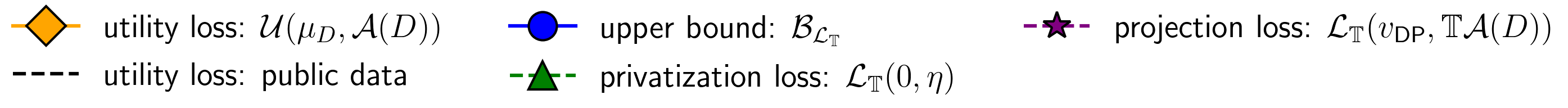}
    \end{subfigure}
    \hspace{0.05\textwidth}
    \begin{subfigure}{0.45\textwidth}
\includegraphics[width=\textwidth]{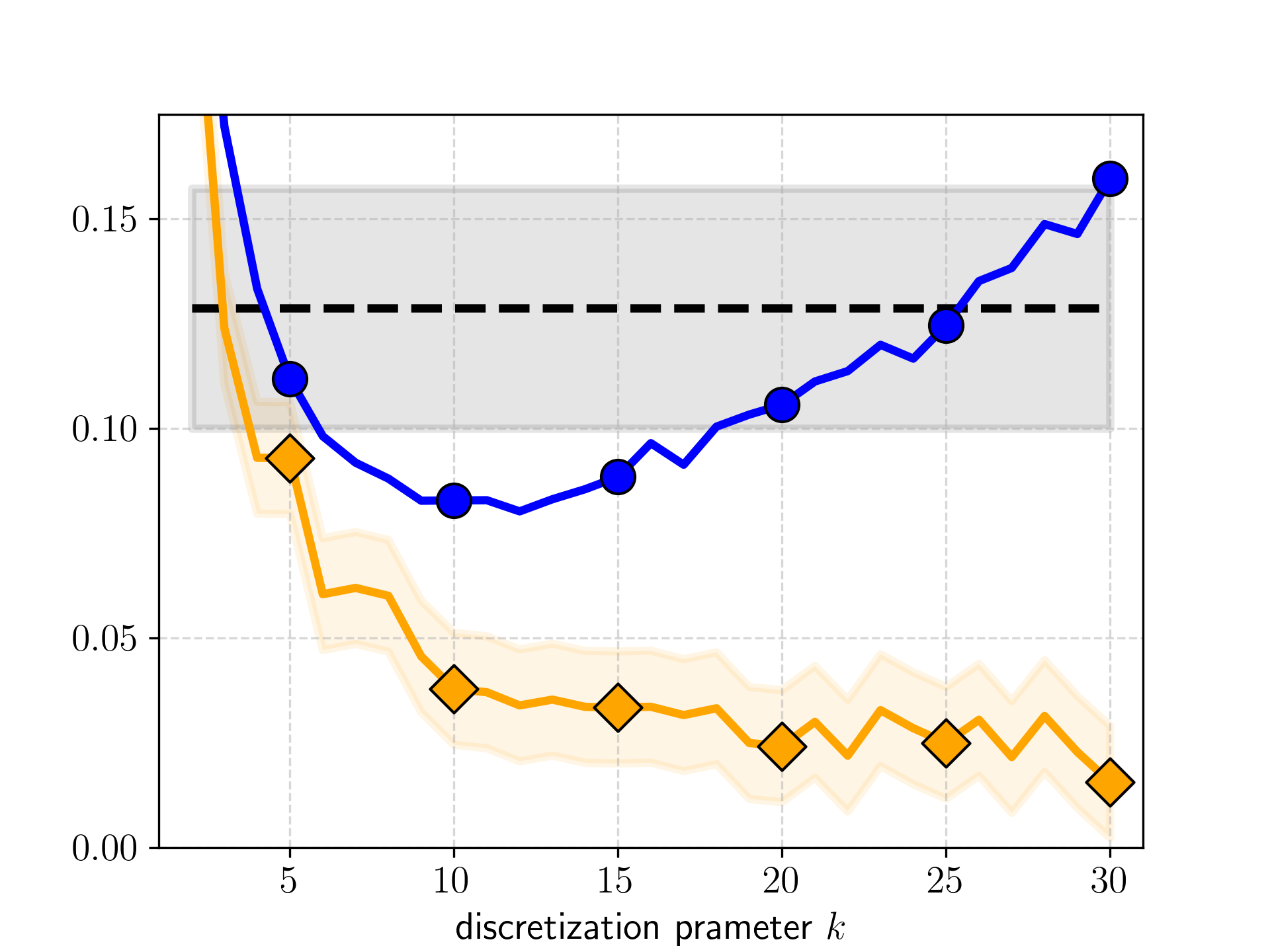}
             \caption{}
         \label{fig:ublt}
    \end{subfigure}
        \begin{subfigure}{0.45\textwidth}
    \centering
\includegraphics[width=\textwidth]{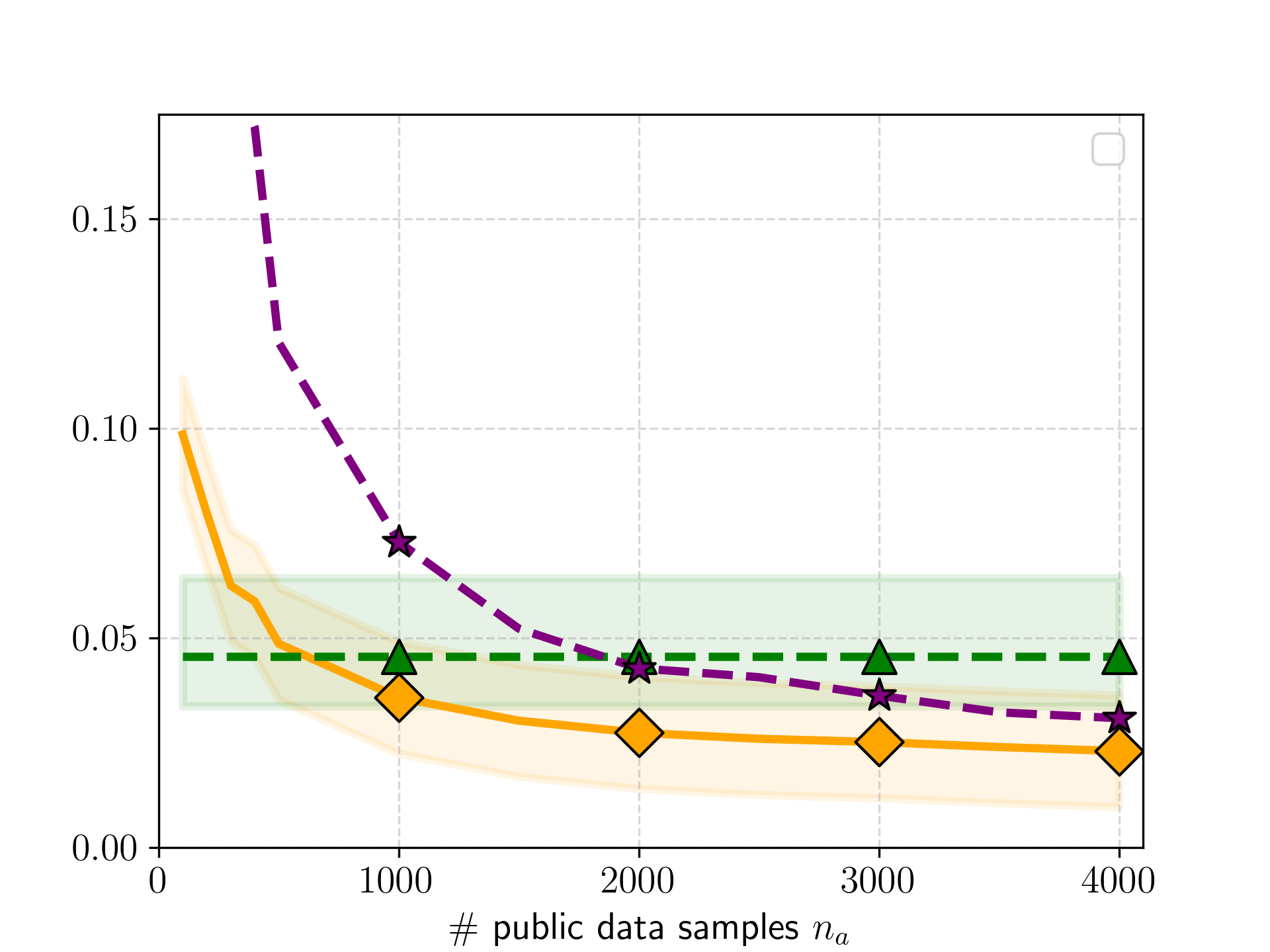}
             \caption{}
         \label{fig:npublic}
    \end{subfigure}
    \caption{\small 
   a)  upper and lower bound of the utility loss (see Appendix~\ref{apx:experiments}) and the utility loss of the empirical measure of the 'raw" public data as a function of 
    function of the discretization parameter $k$ (dashed horizontal line). 
    Moreover, the upper bound from Equation~\eqref{eq:pluginub} with $\delta=0.1$ for Algorithm~\ref{alg:mainalgo} (blue line). b) we compare the term $\utilitylossT(\nudp,\mathbb T \algoD)$  (dashed purple line) with the term 
$\utilitylossT( \nudp, \mathbb T \mu_D) = \utilitylossT(0, \eta) $, (dashed green line) as a function of the size of the public data set. We plot the mean, 0.05 and 0.95 quantiles of  $\utilitylossT(\eta, 0)$  (horizontal lines). Moreover, we plot the upper and lower bounds for the utility loss (yellow line) (see Appendix~\ref{apx:experiments}). 
We choose the discretization parameter $k=25$ for the query operator $\mathbb T$.
}     
    \label{fig:0}
\end{figure}

\subsection{Experimental setting}
For all experiments we use $\epsilon =1$. We use the real-world data sets \textit{ACSIncome}  and \textit{ACSTravelTime} \citep{ding2021retiring} collected from the ''American Community Survey"
and rescale them such that the data lives in the hypercube. For the public data, we use the $n=195665$  samples from California from 2018 (ACSIncome) and the $n=91200$ samples from New York from 2018 (ACSTravelTime). As public data sets, we randomly choose $\pubm = 4000$
samples from Alabama from 2018 and from California from 2014 (ACSIncome) and from Massachusetts from 2018 and New York from 2014 (ACSTravelTime). 
Furthermore, from the ACSIncome data set we only consider the $d=5$  features "AGEP, SCHL, OCCP, POBP, WKHP"  and from the ACSTravelTime data set the $d=4$ features "AGEP, SCHL, PUMA, POVPIP"
(see \citep{ding2021retiring} for the documentation).  If not further specified, we choose the ACSIncome data set with the samples from Alabama from 2018 as public data and the samples form California from 2018 as private data set.

Moreover, while exactly computing the utility loss \eqref{eq:utilityloss} turns out to be computationally infeasible for our choices of $n$ and $\pubm$, we can compute sharp upper and lower bounds as described in  Appendix~\ref{apx:experiments}. 
For all plots, we take the average over 10-independent runs and use 200 samples to approximating the quantiles in Equation~\eqref{eq:ub}.

\subsection{Numerical evaluation using real world data}
\label{sec:experiments}

In this section we present numerical experiments for Algorithm~\ref{alg:mainalgo}. We illustrate in Figure~\ref{fig:1} the certificate from Equation~\eqref{eq:pluginub} and the utility loss for the measure generated by Algorithm~\ref{alg:mainalgo} as a function of the discretization parameter $k$. As we can see, we achieve a significantly smaller utility loss  than when simply using the ''un-optimized" public data. 
Moreover, the certificate from Algorithm~\ref{alg:mainalgo} captures the trend of the utility loss in the beginning and yields a non-trivial guarantee. We refer to Section~\ref{subsec:tightcertificate} for a tighter choice for the certificate.

Moreover, we argue based on Figure~\ref{fig:npublic} that we can measure the ''sub-optimality" of an approximate solution $\mu_{\text{approx}}$ for Step 3 in Algorithm~\ref{alg:mainalgo}
by comparing the terms $\utilitylossT(\nudp,\mathbb T \mu_{\text{approx}})$ and
$\utilitylossT( \nudp, \mathbb T \mu_D) = \utilitylossT(0, \eta) $ (see Figure~\ref{fig:npublic}).
Intuitively,  these two terms capture the ''distances"  of the DP measure 
$\mu_{\text{approx}}$ and the private measure $\mu_D$ to the ''reference point" $\nudp$. Thus, once 
$\utilitylossT(\nudp,\mathbb T\mu_{\text{approx}}) \approx \utilitylossT(\nudp, \mathbb T \mu_D)$, $\mu_D$  and $\mu_{\text{approx}}$ have the same ''distance" to $\nudp$, and therefore we no longer expect an improvement in the utility loss when further minimizing $\utilitylossT( \nudp, \mathbb T \mu_{\text{approx}})$. 
We illustrate this in Figure~\ref{fig:npublic}, where we plot the utility loss and the two terms as a function of the amount of public data samples. By increasing the amount of public data samples, we can improve our approximation of Step 3 in Algorithm~\ref{alg:mainalgo}. As we can see, once $\utilitylossT(\nudp,\mathbb T\mu_{\text{approx}}) \approx \utilitylossT(\nudp, \mathbb T \mu_D)$ the utility stagnates, meaning that we do not benefit from further optimizing $\utilitylossT(\nudp,\mathbb T\mu_{\text{approx}})$ by increasing the amount of public samples.


\begin{figure}[b!]
    \centering
        \begin{subfigure}{\textwidth}
        \centering
\includegraphics[width=\textwidth]{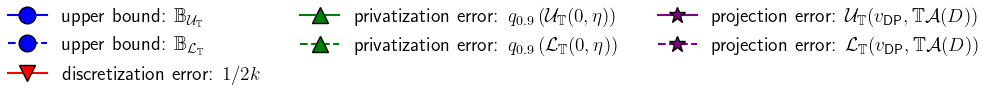}
    \end{subfigure}
    \hspace{0.05\textwidth}
           \begin{subfigure}{0.45\textwidth}
        \centering
\includegraphics[width=\textwidth]{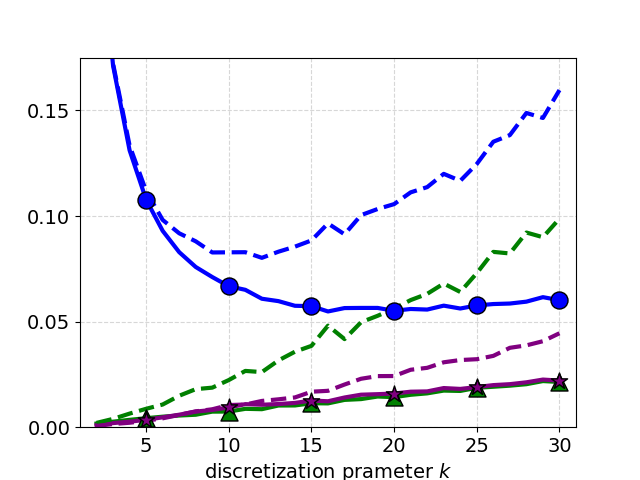}
             \caption{
            }
         \label{fig:2b}
        \end{subfigure}
    \begin{subfigure}{0.45\textwidth}
    \centering
\includegraphics[width=\textwidth]{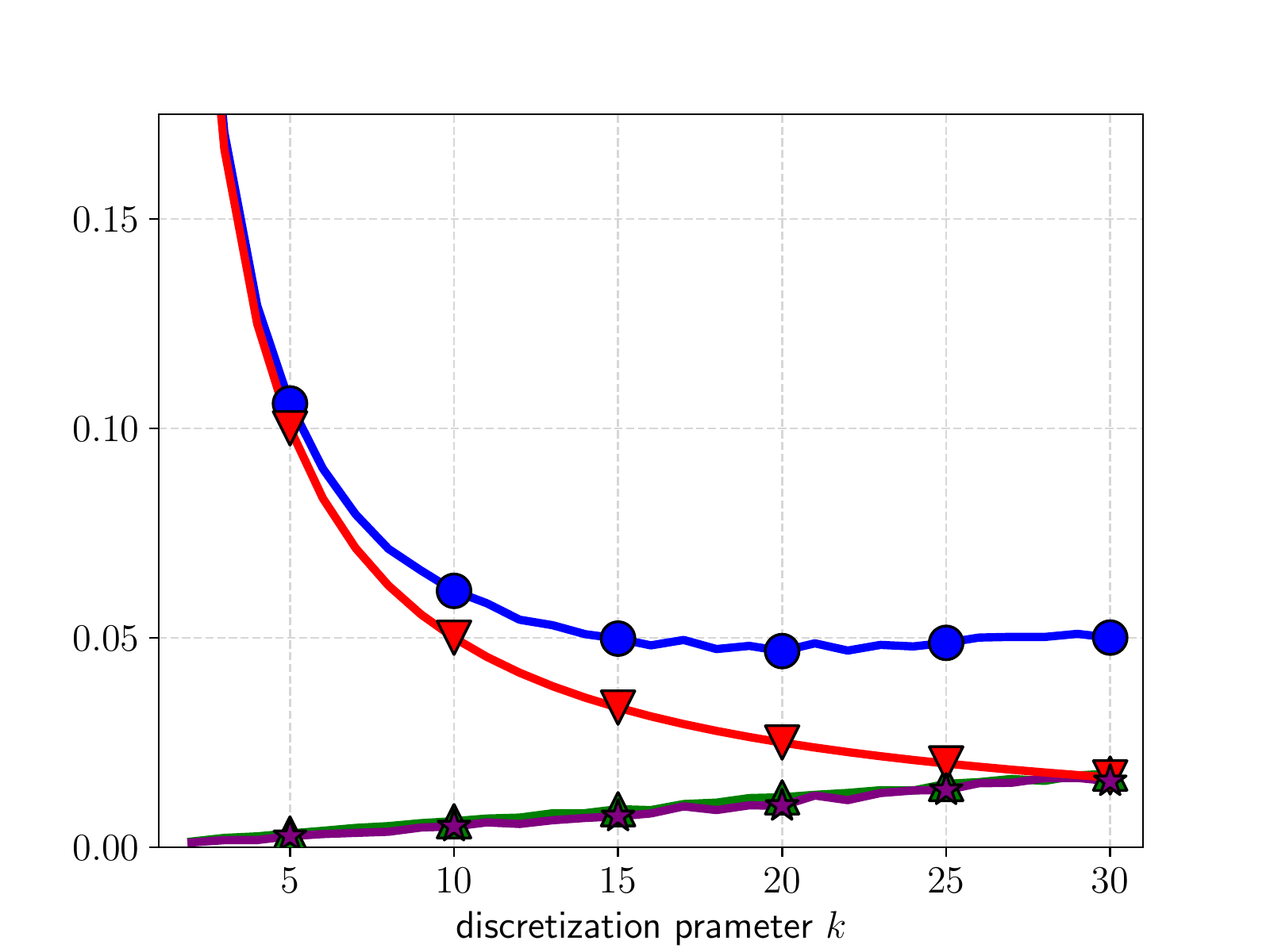}
             \caption{}
         \label{fig:2a}
    \end{subfigure}
    \caption{\small  
(a) the upper bound (solid lines) form Equation~\eqref{eq:pluginub} when using $\proxyutilitytemp = \utilityloss_{\mathbb T}$  instead of $\proxyutilitytemp = \utilitylossT$ (dashed line) for the upper bound in Step 4 in Algorithm~\ref{alg:mainalgo}.
(b) illustration of  the  upper bound from Equation~\eqref{eq:pluginub} and its individual terms for $\proxyutilitytemp = \proxyutility$ with $\delta=0.1$ for Algorithm~\ref{alg:mainalgo} as a function of the discretization parameter $k$. 
}     
    \label{fig:1_new}
\end{figure}

\subsection{A tighter certificate}
\label{subsec:tightcertificate}
In this section we present a tighter choice for the certificate in Step 4 in Algorithm~\ref{alg:mainalgo}. Moreover, we present numerical simulations showing the tightness of the certificate.

\paragraph{A tighter choice  for the proxy utility loss $\proxyutilitytemp = \proxyutility$}
We now present an alternative choice for a proxy utility loss $\proxyutility$ instead of $\utilitylossT$ which allows us to obtain a sharper certificate in Step 4 in Algorithm~\ref{alg:mainalgo}. 
A natural idea to construct a tighter proxy utility loss is to simply "extend" the definition of the utility loss $\utilityloss$ to signed measures on the marginals. We do this as follows:
 for any two vectors $u, v \in \mathbb R ^\Tdim$, we then define the proxy utility loss
\begin{equation}
\label{eq:lossmeasure}
 \utilityloss_{\mathbb T}( v,  u) :=  \max_{S \subset [ d], \vert S \vert =s  } \sup_{\substack{f \in \mathcal F(T_{k,s})\\ \|f \|_{\infty} \leq 1}} \left \vert \sum_{z_i \in T_{k,s}} f(z_i) ~ \left( \omega^{S}_{v}(\{z_i\}) - ~\omega^{S}_{u}(\{z_i\}) \right) \right \vert, 
\end{equation}
where $\mathcal F(T_{k,s})$ is the set of all $1$-Lipschitz continuous functions  over $T_{k,s}$ w.r.t.~the $\ell_\infty$-metric and $\omega^{S}_{v}$ are as in Section~\ref{sec:algo_instantioation}.

Clearly, $\proxyutility$ coincides with $\utilityloss$ on $\image{\mathbb T}$ (i.e.,  satisfies Equation~\eqref{eq:UT}), is jointly translation invariant and satisfies the triangular inequality. Hence,  $\utilityloss_{\mathbb T}$  from Equation~\eqref{eq:lossmeasure} is a valid choice for the upper bound in Equation~\eqref{eq:pluginub} to hold. Concerning the computational complexity, we note that the  RHS in Equation~\eqref{eq:lossmeasure} can be  computed by taking the maximum over ${ d \choose s}$-solutions of linear programs, which has a total run-time complexity 
$O\left({ d \choose s} \text{poly}(k^s) \right)$. 
While this allows for a tractable computation of the certificate
$\ub_{\proxyutility}$ by taking the maximum over linear programs, \emph{optimizing} over the set of measures $\probset$ is non-tractable.
\footnote{The only exception here is the case where $s=d$, in which case we would obtain the algorithm in \citep{Yiyun23}. }
Thus, we keep $\proxyutilitytemp = \utilitylossT$ in Step 3 in Algorithm~\ref{alg:mainalgo} but release in Step 4 in Algorithm~\ref{alg:mainalgo} to certificate $\proxyutilitytemp = \proxyutility$.


\paragraph{Numerical evaluation}
In Figure~\ref{fig:2b} we compare the certificates and the individual terms in Equation~\eqref{eq:ub} for the choices  $\proxyutilitytemp = \proxyutility$ and $\proxyutilitytemp = \utilitylossT$.  We can clearly see that the choice $\ub_{\proxyutility}$ yields a significantly tighter certificate  than the choice $\ub_{\utilitylossT}$, and especially for large choices of $k$.  Moreover, we illustrate in Figures~\ref{fig:2a}  the terms of the certificate from Equation~\eqref{eq:pluginub} for the choice $\proxyutilitytemp = \proxyutility$.  As we can see, increasing $k$ leads to a higher privacy and projection error, but lowers the dicretization error (see Equation~\eqref{eq:pluginub}). This highlights the trade-off between more fine-grained discretizations and the need to add sufficient noise in order to preserve privacy.

\begin{figure}[t!]

    \centering
    \begin{subfigure}{\textwidth}
        \centering
\includegraphics[width=\textwidth]{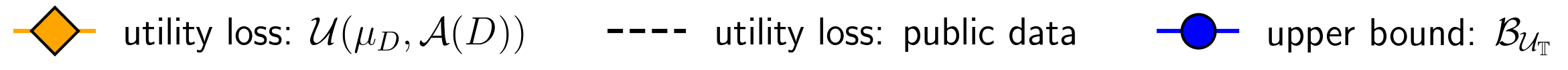}
    \end{subfigure}
    \hspace{0.05\textwidth}
    \begin{subfigure}{0.24\textwidth}
\includegraphics[width=\textwidth]{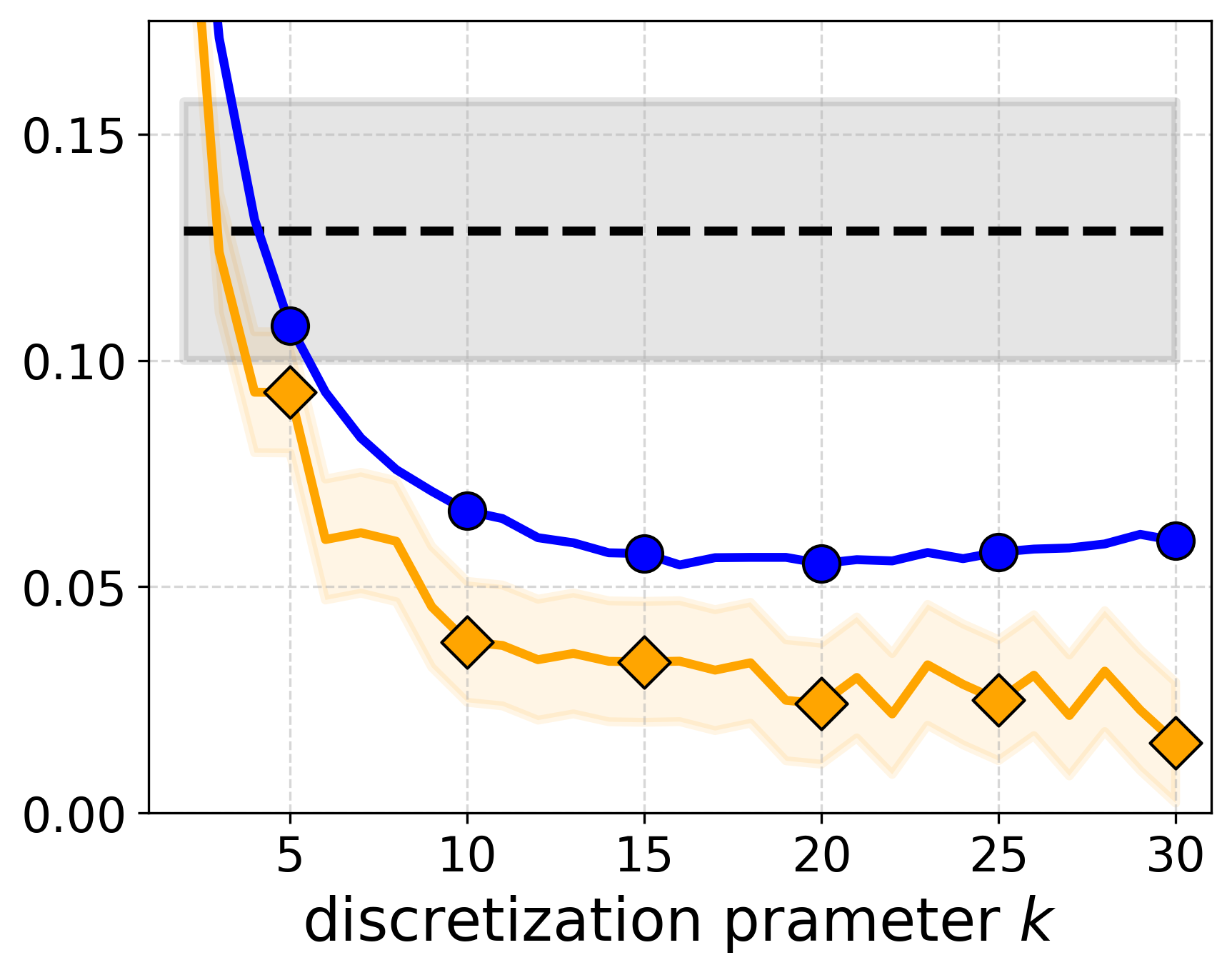}
             \caption{Income AL 18'}
         \label{fig:ub}
    \end{subfigure}
        \begin{subfigure}{0.24\textwidth}
    \centering
\includegraphics[width=\textwidth]{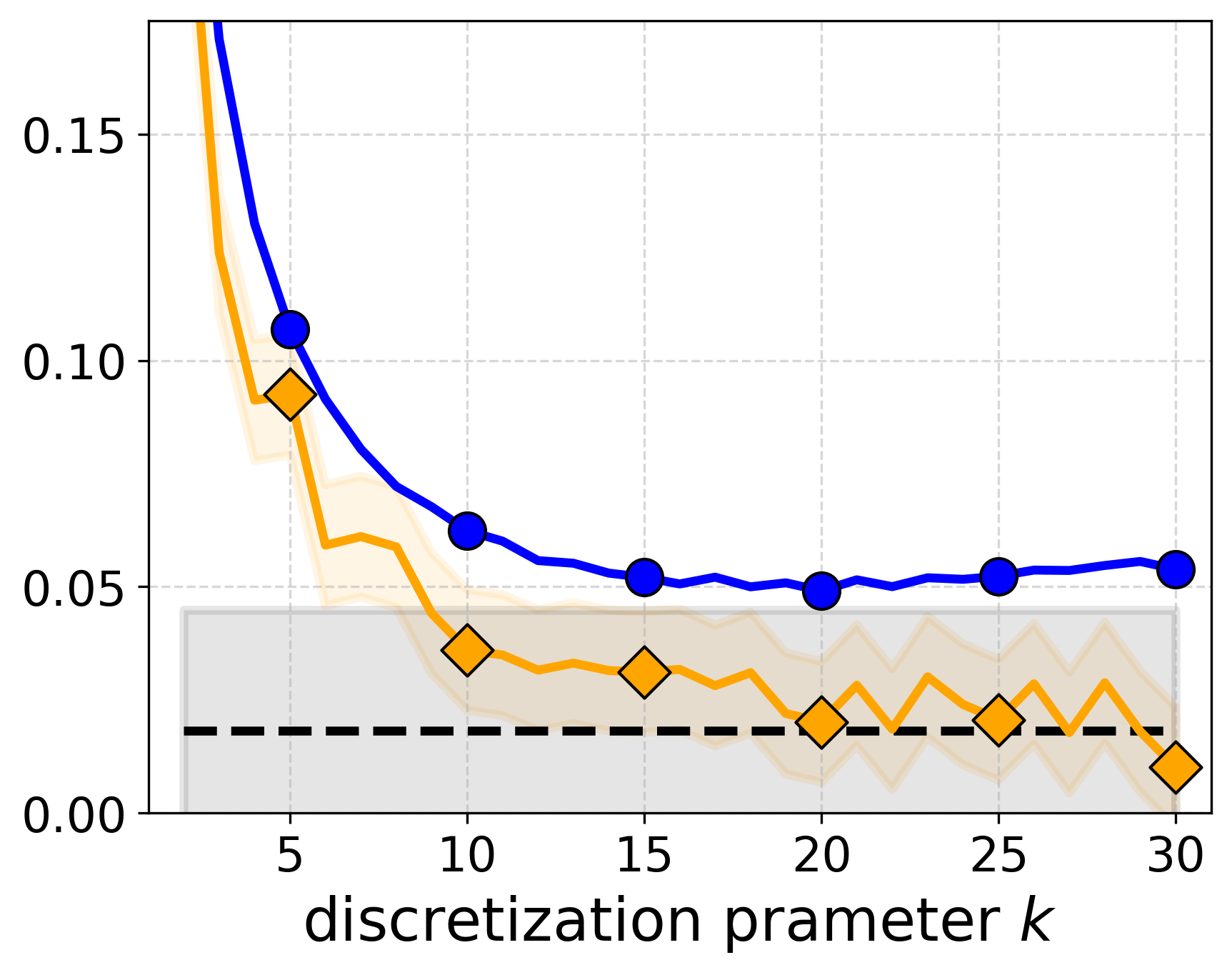}
             \caption{Income CA 14'}
         \label{fig:ub_decomp}
    \end{subfigure}
    \begin{subfigure}{0.24\textwidth}
        \centering
\includegraphics[width=\textwidth]{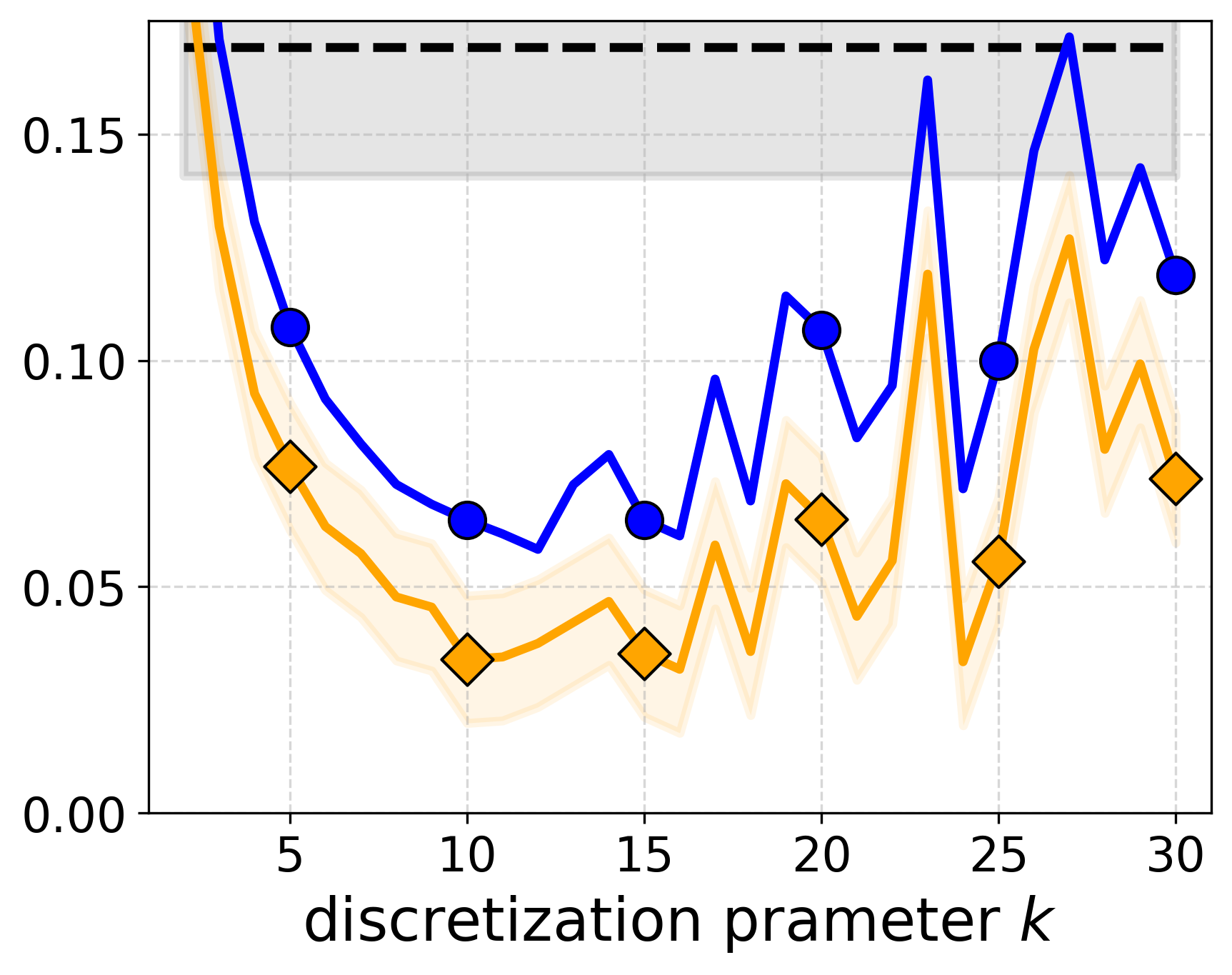}
             \caption{TravelTime MA 18'
            }
         \label{fig:ub_comp}
        \end{subfigure}~
    \begin{subfigure}{0.24\textwidth}
    \centering    \includegraphics[width=\textwidth]{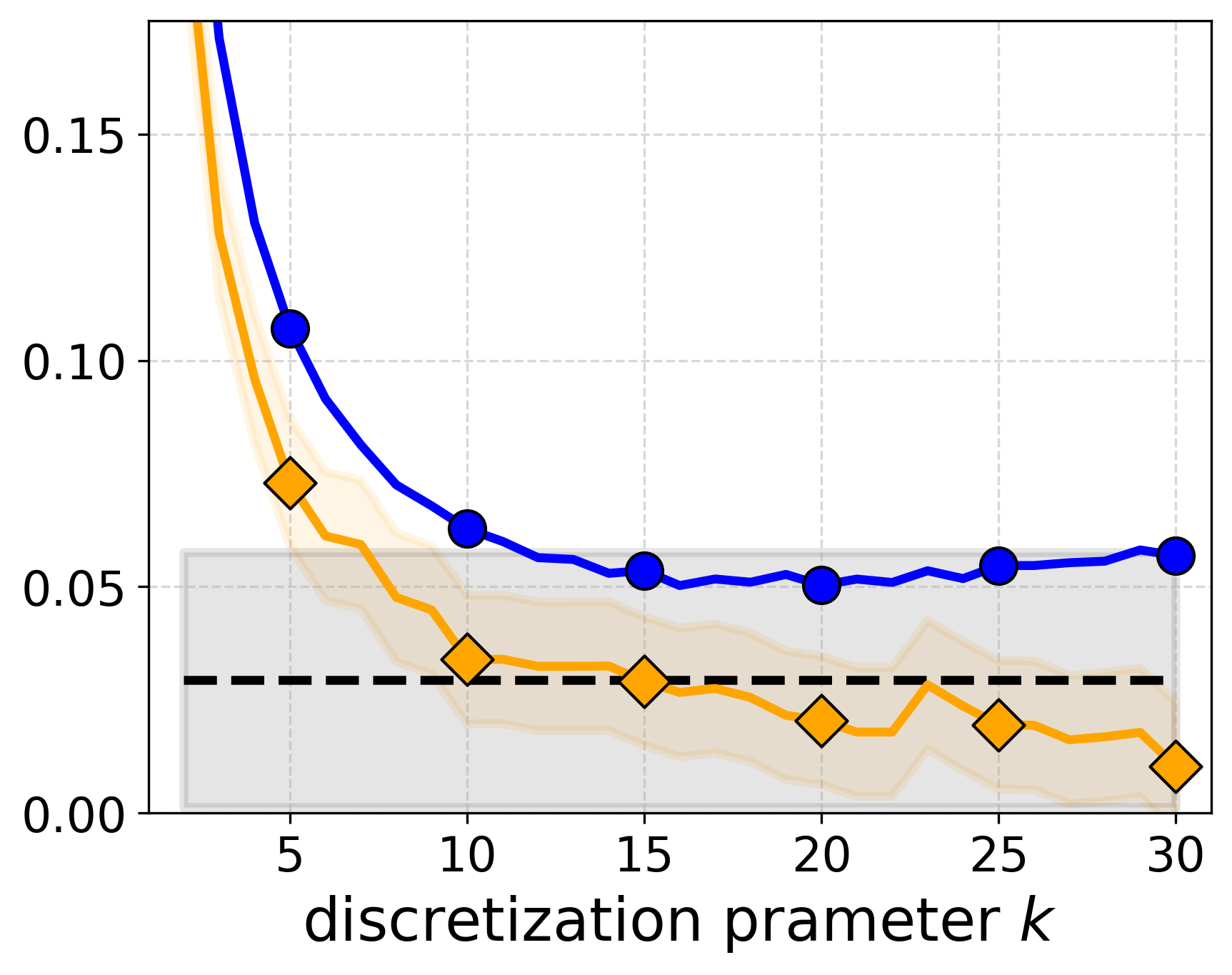}
    \caption{TravelTime NY 14'}
    \end{subfigure}
    \caption{\small 
same curves as in Figure~\ref{fig:ublt} but for the certificate with $\proxyutilitytemp = \proxyutility$ for different combinations of public and private data sets. 
}     
    \label{fig:1}
\end{figure}

Finally, we illustrate in Figure~\ref{fig:1} the certificate from Equation~\eqref{eq:pluginub} and the utility loss for the measure generated by Algorithm~\ref{alg:mainalgo} as a function of the discretization parameter $k$ for different choices of the public and private data set. 
As we can see, when the public data has a large distribution shift,  we achieve a significanly smaller utility loss  than when simply using the ''un-optimized" public data. Moreover, in all plots, the certificate closely matches  the utility loss and correctly captures the trend. 



\section{Related work}
\label{sec:relatedwork}
Releasing datasets privately while minimizing the utility loss for a specific function class is a major challenge in differential privacy. Most studies have focused on preserving utility for counting queries (referred to as statistical queries in~\cite{kearns1998efficient}). The field started with~\cite{blum2013learning}, who applied the Exponential Mechanism algorithm of~\cite{mcsherry2007mechanism} to release a private data set while ensuring that the loss of utility for any a priori known set of counting queries grows at most logarithmically with the set size. Subsequent works by~\cite{dwork2009complexity,mwem,hardt2010multiplicative,roth2010median} 
improved both the statistical and computational complexity and studied fundamental statistical-computational tradeoffs and gaps ~\citep{dwork2009complexity,ullman2011pcps}

In order to reduce sample complexity, several works have considered sparsity assumptions in various ways. 
For example, \cite{blum2013fast} propose an efficient algorithm for queries that take on a non-zero value only on a small subset of an unstructured discrete domain. More related to us, a special class of linear statistical or counting queries are k-way marginals~\citep{dwork2015efficient,liu2021iterative,thaler2012faster}. A k-way marginal query involves fixing the values of \(k\) indices and determining the proportion of data that matches those values.
Further, a range of works~\citep{Barak07,cheraghchi2012submodular,gupta2011privately} also study the query class of k-way conjunctions and provide fast algorithms when \(k\) is small. Further common problems in the privacy literature related to this paper include histogram release 
\citep{abowd2019census,acs2012differentially,hay2009boosting,meng2017different,nelson2019sok,qardaji2013understanding,xiao2010differential,xu2013differentially,zhang2016privtree} and private clustering \citep{balcan2017differentially,ghazi2020differentially,Stemmer20,su2016differentially}.

Finally, recent works ~\citep{boedihardjo2022private,Yiyun23, wang2016differentially} studied the case where $\mathcal F$ is the class of
all $1$-Lipschitz continuous functions,
resulting in the utility
loss~\eqref{eq:utilityloss} equaling the $1$-Wassterstein
distance.
A small Wasserstein distance is desirable in many practical
applications as it for instance guarantees that clusters present in the original data remain
preserved (see the discussion in \citep{boedihardjo2022private}). 
However, prior results \citep{boedihardjo2022private, Yiyun23} also suggest that ensuring a small Wasserstein distance requires exponentially many samples in the dimension.
For example, for the $d$-dimensional hypercube~$[0,1]^d$ (with $d \geq 2$) equipped with the
$\ell_{\infty}$-metric, the papers \citep{boedihardjo2022private, Yiyun23} together show that the optimal expected  utility loss is of order 
\begin{equation}
\label{eq:ubfull}
   \mathbb E ~ \utilityloss(\mu_D, \algoD) \asymp  \left( \frac{1}{n \epsilon}\right)^{1/d},
\end{equation}
where $n$ is the size of the data set $D$ and $\mu_D$ its corresponding empirical measure.


\section{Conclusion and future work}
\label{sec:conclusion}
Especially in sensitive domains, we desire a certificate for the maximum utility loss  to ensure that the data is provably minimally affected by the DP mechanism. We take a step in this direction by introducing Algorithm~\ref{alg:mainalgo} in Section~\ref{sec:algo}, which simultaneously releases a DP discrete probability measure and provides a certificate for the maximum utility loss when $\mathcal{F}$ is the class of all $s$-sparse Lipschitz continuous functions. As shown in Section~\ref{sec:mainthm}, our algorithm achieves an optimal non-asymptotic exponential decay rate for the expected utility loss and effectively overcomes the curse of dimensionality for moderate choices of $s$.

\paragraph{Future work}
 The certificate in Algorithm~\ref{alg:mainalgo} can be computed for any "approximate" solution and we leave practically meaningful,  efficient approximations of  Step 3 in Algorithm~\ref{alg:mainalgo} as  future work. Moreover, we motivate theoretical research on the right dependency on $d$ in Theorem~\ref{cor:rates}. Improving the upper bound would likely result in a novel algorithm with potentially practical applications, while an improved lower bound would require the development of new mathematical ideas and provide evidence for the optimality of Algorithm~\ref{alg:mainalgo}.

\newpage

\bibliography{bib}

\appendix

\section{Experimental Setting}
\label{apx:experiments}


For completeness, we describe the upper and lower bounds for the utility loss \eqref{eq:wassersteinsparse} used in Figures~\ref{fig:0} and \ref{fig:1}.
Although we can compute the Wasserstein distance for discrete measures using linear programming, for the choices of $n$ and $\pubm$ considered in this paper, a direct computation turns out to be computationally infeasible. Instead, for any measures $\mu, \mu'$, we can bound the wasserstein distance using the triangle inequality 
\begin{align}
    \utilityloss(\mu, \mu' ) &\leq  \utilityloss( P^{T_{k,d}}_{\#} \mu,  P^{T_{k,d}}_{\#} \mu' ) + \utilityloss(\mu, P^{T_{k,d}}_{\#}  \mu ) +    \utilityloss( P^{T_{k,d}}_{\#} \mu', \mu' ) ~~\text{and} \label{eq:ubdist}\\
 \utilityloss(\mu, \mu' ) &\geq  \utilityloss( P^{T_{k,d}}_{\#} \mu, P^{T_{k,d}}_{\#} \mu' ) - \utilityloss(\mu,P^{T_{k,d}}_{\#}  \mu ) -  \utilityloss( P^{T_{k,d}}_{\#} \mu', \mu' )\label{eq:lbdist}, 
\end{align}
where $P^{T_{k,d}} : \probset \to \probsetk$ projects the space $T$ to $T_{k,d}$.  We remark that $\utilityloss( P^{T_{k,d}}_{\#} \mu,  P^{T_{k,d}}_{\#} \mu' ) = \utilityloss(  \Tinv \mathbb T\mu,    \Tinv \mathbb T\mu' ) = \proxyutility(T\mu,   \mathbb T\mu')$, which can be computed efficiently (for moderate choices of $k$) as described in Section~\ref{subsec:algo}. Moreover, by construction  $\utilityloss( P^{T_{k,d}}_{\#} \algoD, \algoD ) = 0$, and  for any data set $ D$, we can bound
\begin{equation}
    \utilityloss(P^{T_{k,d}}_{\#}\mu_D, \mu_D ) \leq \max_{S \subset  [ d]; \vert S \vert =s} \frac{1}{n}\sum_{i=1}^{\vert D\vert} \| P^S z_i^S - P^{T_{k,d}}_{\#} P^S z_i\|_\infty,
\end{equation}
Finally, we plot in Figures~\ref{fig:0} and \ref{fig:1} the upper and lower bounds from Equation~\eqref{eq:ubdist} and \ref{eq:lbdist}, as well as the term $\utilityloss( P^{T_{k,d}}_{\#}\mu_D, P^{T_{k,d}}_{\#}\algoD ) $ and $\utilityloss(P^{T_{k,d}}_{\#}\mu_D,P^{T_{k,d}}_{\#}\mu_{\pubdata} ) $, respectively, for $k =30$.

\section{Haar Basis}\label{sec:haar}
\label{apx:haar}
In this section, we give a quick introduction to the Haar matrices, which play a crucial role in the proofs= of Theorem~\ref{thm:mainthm}. For a more in depth discussion please refer to \citep{haar}. 

\begin{figure}[H]
    \centering
        \includegraphics{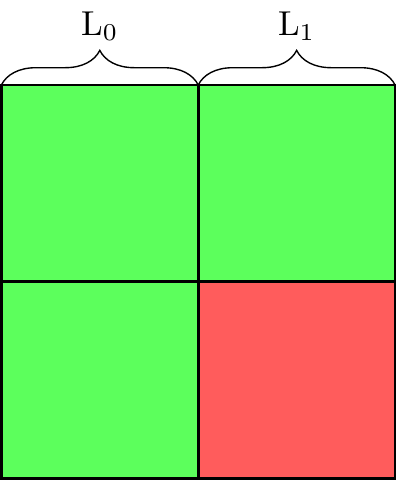}~~
        \includegraphics{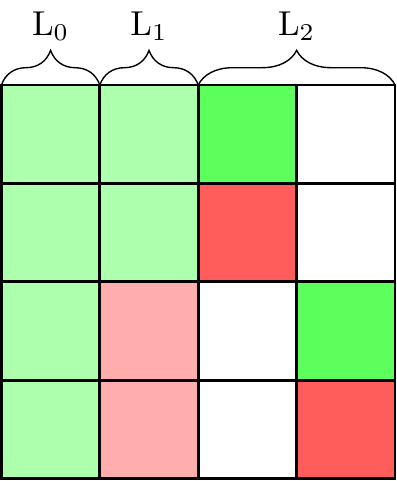}~~
        \includegraphics{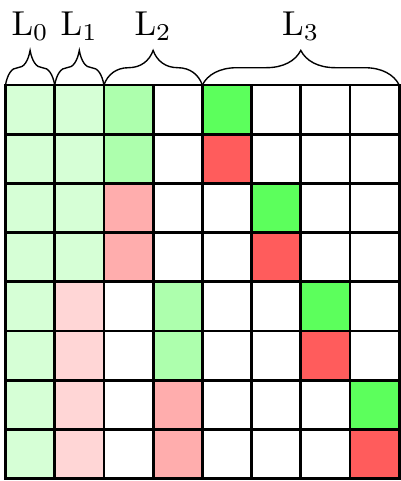}
    \caption{Evolution of the transposed Haar basis shwoing $M_1$, $M_2$ and $M_3$. Green cells contain positive values, red cells negative values while white cells contain 0. The intensity of a cell correspond to the magnitude of the value within it.}
    \label{fig:haar}
\end{figure}

\begin{figure}[H]
    \centering
        \centering
        \includegraphics{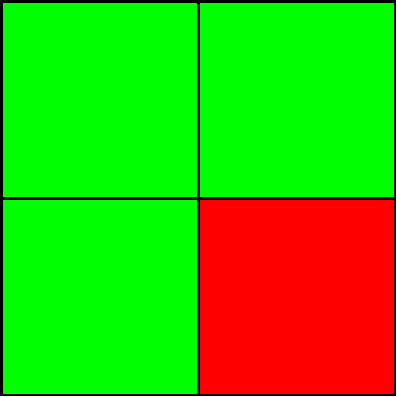}~~
        \includegraphics{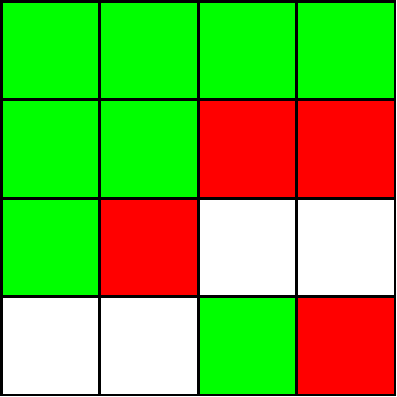}~~
        \includegraphics{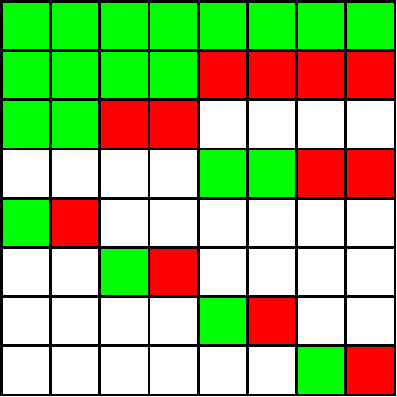}
        \label{fig:h3}
    \caption{Evolution of the inverse of the transposed Haar basis showing $M^{-1}_1, M^{-1}_2$ and $M^{-1}_3$. Green cells contain value 1, red cells -1 while white cells contain 0.}
    \label{fig:haarinv}
\end{figure}

The $k$-th transposed Haar matrix $M_k$ is a $2^k \times 2^k$ matrix. We can separate the columns into $k + 1$ levels $L_0, \ldots, L_k$ where  level $L_l$ contains $\max\{1, 2^{l - 1}\}$ columns (see \Cref{fig:haar}). The level $L_0$ only consists of the first column, the level $L_1$ contains the next column, and the level $L_2$ the following two columns and so on. Moreover, the absolute values of all non-zero elements in the level $L_l$ in $M_k$ are all equal to $\max\{1, 2^{l - 1}\} / 2^k$. Furthermore, a key property of transposed Haar matrices is that the columns are sparse; each column in the level $L_l$ in $M_k$ contains exactly $2^{\min\{k, k - l + 1 \}}$ non-zero elements.
We visualize in Figure~\ref{fig:haar} the transposed Haar matrices $M_1$, $M_2$ and $M_3$. The pattern can be extended to general~$M_k$.
It is straight forward to verify that all the columns in any $M_k$ are orthogonal. Consider any two columns in the same level; their support is disjoint and their scalar product vanishes.  On the other hand, for any two columns in different levels, their support is either disjoint or the support of one is contained in an index set where the values in the other column is constant.  Hence, as every column has an equal number of positive and negative values with equal magnitude, we can conclude that either way the scalar product is zero. Consequently, if we scale the columns appropriately, the Haar basis matrix would be orthogonal. $M_k^{-1}$ is therefore equal to $M_k^T$ with the columns scaled appropriately. Finally, we note that the appropriate scaling is such that all non-zero elements of the inverse have absolute value 1, as visualized in Figure~\ref{fig:haarinv}.


We can now state the following lemma which we use in the proof of  Theorem~\ref{thm:mainthm}. These properties of the transposed Haar matrix have already been implicitly used in the proofs in \citep{boedihardjo2022private}.

\begin{lemma} 
\label{lm:haar_properties}
For any $m\geq 0$ and $\Phi = (\ceil{\log_2(\Tdim)} + 1) M_{\ceil{\log_2(\Tdim)}}$ with $M_k$ the $k$-th transposed Haar Matrix, it holds that $ \| \Phi^{-1} \|_1 \le 1$ and
    \begin{equation}
   \max_{i\in[ \Tdim]} \left \| \sum_{j = 1}^i \Phi_j \right \|_1 \le (\ceil{\log_2(\Tdim)} + 1)^2 \quad \text{and} \quad
       \max_{i\in[\Tdim]} \left \| \sum_{j = 1}^i \Phi_j \right \|_2 \le (\ceil{\log_2(\Tdim)} + 1)^{3/2}
    \end{equation}
\end{lemma}
\paragraph{Proof of Lemma~\ref{lm:haar_properties}}

\begin{itemize}
    \item For the first property, we note that by definition, $\| \Phi^{-1} \|_1 = \max_{l} \| \Phi^{-1}_l \|_1$ where $\Phi^{-1}_l$ is the $l$-th column of $\Phi^{-1}$. By the construction (see Figure~\ref{fig:haarinv}), we have that for all $k$, $\|(M^{-1}_k)_l\|_1 = k + 1$, and thus  $\|  M^{-1}_k \|_1 = k + 1$. Therefore, we get that $\| \Phi^{-1} \|_1 = \| M^{-1}_k \|_1 / (k + 1) = 1$.

    \item For the second and third property we notice that due to the disjoint support between the columns within a level, the contiguous support of each column and the equal number of positive and negative values within a column (see Figure~\ref{fig:haar}), when summing the $j$-first columns of $M_k$ there will be at most one non-zero element in each level. As the 1-norm of each column is 1, we know that the magnitude of this non-zero element is upper bounded by 1. Furthermore, in total we have $k + 1$ levels. Hence, $\| \sum_{l = 1}^i (M_k)_l \|_2 \le \sqrt{k + 1}$ for any $i \in [k + 1]$. Thus, for any $i\in [2^k]$ we get that  $\| \sum_{l = 1}^i \Phi_l \|_2 \le (k + 1)^\frac{3}{2}$ and $\| \sum_{l = 1}^i \Phi_l \|_1 \le (k + 1)^2$.
\end{itemize}


\section{Extension of Theorem~\ref{cor:rates} to  general metric spaces}
\label{sec:apxmainthm}
In this section we generalize the setting in Theorem~\ref{cor:rates}  to general metric spaces  in Theorem~\ref{thm:mainthm}. 
Generally, we can ask for an algorithm $\mathcal A$, taking a data set $D \in T^n$ on some measurable space  $(T, \mathcal B)$ as input, to achieve a small utility loss \eqref{eq:utilityloss} over a function class
\begin{equation}
\label{eq:funcclasssparse}
    \mathcal F = \bigcup_{i=1}^K \mathcal F^{(i)} :=  \bigcup_{i=1}^K\{f^{(i)}\circ \projectfunci\vert f^{(i)}~\text{is 1-Lipschitz~w.r.t.~} \rho^{(i)}\},
\end{equation}
with $(T^{(i)}, \rho^{(i)})_{i=1}^K$ being some set of metric spaces and surjective measurable functions $\projectfunci : T \to T_{(i)}$ serving as ``projections'' of $T$ to $T^{(i)}$. We assume that the push-forward $\sigma$-algebras generated by $h^{(i)}$ coincide with the $\sigma$-algebras generated by $\rho^{(i)}$ on $T^{(i)}$.

Analogous to Equation~\eqref{eq:wassersteinsparse}, by the
Kantorovich-Rubinstein Duality Theorem~\citep{Villani2008OptimalTO},
the utility loss is exactly the
maximum of the Wasserstein distances w.r.t. the metrics $\rho_{(i)}$,
\begin{equation}
\label{eq:Wassersteinsparse}
    \utilityloss(\mu,  \mu') = \max_{i\in [K]} W_1(\projectopi \mu,  \projectopi  \mu').
\end{equation}
To give an example, in the context of Theorem~\ref{cor:rates}, we have $K ={ d \choose s}$ and $T^{(i)} = [0,1]^s$  are the $s$-sparse marginals of $T$ with $\rho^{(i)}(x,y) = \|x-y\|_{\infty}$. Furthermore, $h^{(i)}$ are the functions projecting  $x\in [0,1]^d$ to the corresponding $s$-dimensional subspaces and $\projectopi \mu$ are the marginal measures of $\mu$ on the corresponding $s$-dimensional subspaces.

We now turn to the main result of this section, Theorem~\ref{thm:mainthm}, which
provides a general upper bound for the utility loss over the function
class $\mathcal F$. When $K=1$ we can directly obtain the upper bounds
in Theorem~\ref{thm:mainthm} from the results in
\citep{boedihardjo2022private}. The main technical contribution of this section
is to show that the price to pay if $K>1$ is at most linear in $K$
(plus a logarithmic factor). 
Let $N(\T, \rho,\discdelta)$ be the covering
 number of $\T$, we have:

\begin{theorem}\label{thm:mainthm}
For the setting described above, there exists a
universal constant $c >0$ and a randomized algorithm $\gA$ that takes
a data set  $D \in T^n$ of size $n$  as input and returns a
finitely-supported measure $\algoD \in \probset$ such that
$\gA$ is $\epsilon$-DP private and has expected utility
loss~\eqref{eq:utilityloss} over the function class $\mathcal F$,
defined in Equation~\eqref{eq:funcclasssparse}, at most 
\begin{equation} 
\begin{split}
\label{eq:eq1mainthm}
\mathbb E&~\f{\utilityloss}{\mu_D, \f{\gA}{D}} \leq t \\
&+ c  K \max_{i\in [ K]}  \bigg[\frac{\left(\ceil{\log_2\f{N}{T^{(i)},\rho^{(i)},t}} + 1 + \log(K)\right)^2 }{n \epsilon}
\int_{t/2}^{\f{\mathrm{diam}}{T^{(i)}}/2}\f{N}{T^{(i)},\rho^{(i)},x}dx\bigg].
\end{split}
\end{equation}
\end{theorem}

We now divide the proof of Theorem~\ref{thm:mainthm} into two parts. First, we prove in Appendix~\ref{apx:generalproofk1} Theorem~\ref{thm:mainthm} for the known case when $K=1$. While the proofs builds upon the ideas in \cite{boedihardjo2022private}, we present the proof in a different structure based on  Equation~\eqref{eq:utilityUBgeneral} and Definition~\ref{def:utilityloss} from Section~\ref{sec:mainthm}. This structure is crucial since it then allows us in a second part in Section~\ref{apx:generalproofkall} to extend the proofs to the case where $K>2$ 




\subsection{{Proof of Theorem~\ref{thm:mainthm}} when $K=1$}
\label{apx:generalproofk1}
 We first present a proof for the case where $K=1$, where we can assume w.l.o.g.~that $h_1$ is the identity function. In this case, we obtain exactly the results from Section~7 in \citep{boedihardjo2022private}. 
 The proof 
 consists of
 three parts, where we first construct a query operator $\mathbb T$ and a proxy utility loss $\utilitylossT$
 (Definition~\ref{def:utilityloss}).

 \paragraph{Construction of the query operator $\mathbb T$ and the
 right-inverse $\Tinv$:} Let $\Tdim= N(\T, \rho,\discdelta)$ be the covering
 number of $\T$ and let $T_\Tdim$ be the centers of any minimal
 $t$-covering of $\T$. Furthermore, let $\mathbb D_{T_\Tdim} :
 \measure(\T) \to \measure(T_\Tdim)\subset \measure (\T)$ be the
 projection operator which constructs a probability measure on
 $T_\Tdim$ by dividing the space $\T$ into $m$ disjoint measurable
 neighborhoods around the points in $T_\Tdim$ of diameter at most
 $t$.
 Given any indexing of the elements in $T_\Tdim$, we can
 straightforwardly define a bijection from $\measure(T_\Tdim)$ to the
 probability simplex $\mathcal V = \{z \in \mathbb R^\Tdim: \|z\|_1
 =1, z_i \geq 0\}$ and thus complete the construction of the operator
 $\mathbb T$. Furthermore, let the
 right-inverse $\Tinv: \mathcal V \to \measure(T_\Tdim)
 \subset \probset$ be any operator such that~$\Tinv
 \mathbb T = \mathbb D_{T_\Tdim}$. 

To simplify the following analysis, we now describe how to choose a
particular indexing of the elements in $T_\Tdim$ based on the analysis
in~\cite{boedihardjo2022private}.  Proposition 6.5 and Equation (7.4)
in \citep{boedihardjo2022private} together guarantee the existence of
a finite set $\Omega =\{w_1, \cdots, w_\Tdim\} \subset [0,L]$  as well
as a $1$-Lipschitz (w.r.t.~$\rho$) bijection $f: ~\Omega \to T_\Tdim$
with
\begin{equation}
    L =  64 \int_{t/2}^{\diam(T)/2} N(T, \rho, x) dx.
\end{equation}
  Note that w.l.o.g.~we can assume that $w_i \leq w_{i+1}$
 which thereby implicitly induces a Hamiltonian path on $T_\Tdim$ of length at most $L$. We can now define the indexing of the elements in $T_\Tdim$ by $z_i = f(w_i)$, 
  which allows us to  complete the construction of the operator $\mathbb T$.

\paragraph{Step 3 in Algorithm~\ref{alg:mainalgo}: proxy utility loss  $\utilitylossT$:} 
One of the key ideas of~\cite{boedihardjo2022private} is to reduce the the problem of constructing a private measure on $T$ to that of constructing a private measure on an interval on $\mathbb R$. This is done via the bijection $f$ introduced in the previous paragraph. Along these lines, we now show how we can make use of the bijection $f$ to construct a proxy utility loss $\utilitylossT$ satisfying the conditions in Definition~\ref{def:utilityloss}. 

If $\mu_{T_\Tdim}\in\measure(T_\Tdim)$, then let $\mu_{T_\Tdim,f}$ be the push-forward measure of  $\mu_{T_\Tdim} \in \measure(T_\Tdim)$, i.e. the measure on $\measure(\Omega)$ such that for any $A \subset \Omega$,
$\mu_{T_\Tdim,f}(A) = \mu_{T_\Tdim}(f(A))$. Since $f$ is a $1$-Lipschitz continuous function, the Wasserstein distance of any two measures $\mu_{T_\Tdim}, \mu_{T_\Tdim}' \in \measure(T_\Tdim)$ is upper bounded by
\begin{equation}
\begin{split}
\label{eq:ubHfull}
  \utilityloss(\mu_{T_\Tdim}, \mu_{T_\Tdim}') &= W_1(\mu_{T_\Tdim}, \mu_{T_\Tdim}') \leq W_1(\mu_{T_\Tdim,f}, \mu_{T_\Tdim,f}') \\
  &=
      \| F_{\mu_{T_\Tdim,f}} - F_{\mu_{T_\Tdim,f}'}\|_{L^1(\mathbb R)} \\&
      =\sum_{j=1}^\Tdim (w_{j+1} - w_j) \left\vert \sum_{i=1}^j (\mu_{T_\Tdim}(f^{-1}(w_{i})) -  \mu_{T_\Tdim}'(f^{-1}(w_{i})))\right\vert \\&
      =\sum_{j=1}^\Tdim (w_{j+1} - w_j) \left\vert \sum_{i=1}^j (v_i -  v'_i)\right\vert 
     \end{split}
      \end{equation}
where the second equality follows from the identity
in~\citep{vallender74}, $F_{\mu_{T_\Tdim,f}}$ is the cumulative distribution
function of the measure $\mu_{T_\Tdim,f}$, and we use the notation $v
= \mathbb T \mu_{T_\Tdim}$, $v' = \mathbb T \mu_{T_\Tdim}'$, and
$w_{\Tdim+1}= L$.  Using the RHS of~\Cref{eq:ubHfull} we can now define the
utility proxy utility loss function $\utilitylossT$ on $\mathbb R^\Tdim$
\begin{equation}
   \utilitylossT(v, v') :=\sum_{j=1}^\Tdim (w_{j+1} - w_j) \left\vert \sum_{i=1}^j (v_i -  v'_i)\right\vert,
         \label{eq:defofHfull}
\end{equation}
and therefore have
$\utilityloss(\mu_{T_\Tdim},\mu_{T_\Tdim}')\leq\utilitylossT(\mathbb
T\mu_{T_\Tdim},\mathbb T\mu_{T_\Tdim}')$ for all
$\mu_{T_\Tdim},\mu_{T_\Tdim}'\in\measure(T_\Tdim)$.
Hence, we conclude that $\utilitylossT$ satisfies the conditions in Definition~\ref{def:utilityloss}.

\paragraph{Upper bound for the utility loss $\utilityloss$:} Finally, we can prove the result by upper bounding the utility loss using Equation~\eqref{eq:utilityUBgeneral}.
First, the ``projection'' error term in
Equation~\eqref{eq:utilityUBgeneral} can be upper bounded by
 \begin{equation}
     \sup_{\mu \in \probset} \utilityloss(\mu, \Tinv \mathbb T \mu ) =  \sup_{\mu \in \probset} W_1(\mu, \Tinv \mathbb T \mu ) = \sup_{\mu \in \probset}  W_1(\mu, \mathbb D_{T_\Tdim} \mu )\leq \discdelta, 
 \end{equation}
 which is a consequence of the fact that $\mathbb D_{T_\Tdim}$ moves every point mass at most distance  $\discdelta$.  
 
Next, we bound second term in Equation~\eqref{eq:utilityUBgeneral},  $2~\utilitylossT(\mathbb T \mu_D, \nudp)$. To do so, we first need to choose the matrix $\Phi$ in Equation~\eqref{eq:noiseeq}. As first suggested in \citep{xiao2010differential} and also in \citep{boedihardjo2022private}, we can choose $\Phi =(\ceil{\log_2(\Tdim)} + 1) M_{\ceil{\log_2(\Tdim)}}$ where
 $M_k$ is the $k$-th Haar Matrix (see Appendix~\ref{apx:haar}) and then apply the Laplace mechanism as in Lemma~\ref{lm:privacy_phi}. 
 First note that $
 \Delta_{\mathbb T} \leq \frac{2}{n}$ where $\Delta_{\mathbb T}$ is
 defined in Lemma \ref{lm:privacy_phi}. Indeed, the vectors
 $\mathbb T\mu_D$ and $\mathbb T\mu_{D'}$ are representations for the
 discretized measures $\mathbb D_{T_\Tdim} \mu_D$ and $\mathbb
 D_{T_\Tdim} \mu_{D'}$, which differ at most in two points. By Lemma~\ref{lm:privacy_phi} we therefore need to draw 
${\tilde \eta  \sim \left(\text{Lap}\left(\frac{2\|\Phi^{-1}\|_1}{n \epsilon} \right)\right)^{m_{\Phi}}}$. A straightforward calculation then yields the following upper bound:
        \begin{equation}
        \begin{split}
          \mathbb E ~&\utilitylossT(\mathbb T \mu_D, \nudp) =  \mathbb E ~\utilitylossT(\mathbb T \mu_D, \mathbb T \mu_D + [\Phi \tilde \eta]_{1:\Tdim}) \\
          &=\sum_{j=1}^\Tdim (w_{j+1} - w_j) \mathbb E_{\tilde \eta \sim \left(\text{Lap}\left(\frac{2 \|\Phi^{-1}\|_1}{n \epsilon} \right)\right)^{\Tdim_{\Phi}}}\left\vert \sum_{i=1}^j (\Phi \tilde \eta)_i\right\vert  \\
      &\stackrel{(\text{Jensen})}{\le}  L \max_{j = 1}^\Tdim  \mean{ \text{std} \left(\sum_{i = 1}^j (\Phi \tilde \eta)_i \right)}{\tilde \eta \sim \left(\text{Lap}\left(\frac{2 \|\Phi^{-1}\|_1}{n \epsilon} \right)\right)^{\Tdim_{\Phi}}}   = \frac{2\sqrt{2}L \|\Phi^{-1}\|_1}{n \epsilon} \cdot \max_{j = 1}^\Tdim \left \| \sum_{i = 1}^j \Phi_i \right \|_2, \label{eq:marchboundexp}
        \end{split}
        \end{equation}
 where $\Phi_i$ is the i-th row of $\Phi$. 
We then obtain the desired  upper bounds when applying Lemma~\ref{lm:haar_properties} in Appendix~\ref{apx:haar}.

\subsection{Full proof of Theorem~\ref{thm:mainthm} for arbitrary $K>1$}
\label{apx:generalproofkall}

We now show how we can extend the result for the case
where $K=1$~(in Section~\ref{apx:generalproofkall}) to
the case of $K>1$ by changing the query operator $\mathbb T$. We construct \(\mathbb{T}\) with $\Tdim
= \sum_i \Tdim_{(i)}$ with $\Tdim_{(i)} = \f{N}{T^{(i)},\rho^{(i)},t}$, by simply stacking the operators $\mathbb
T_{(i)} \circ \projectopi :\measure(T) \to \mathbb R^{\Tdim_{(i)}}$
where $\mathbb T_{(i)}$ are the query operators described in
Section~\ref{apx:generalproofk1} for the metric spaces $(T_{(i)}, \rho_{(i)})$.
It is then straightforward to verify that for any right inverse, the
``projection" error $\utilityloss(\mu_D, \Tinv \mathbb T
\mu_D)$ from
Equation~\eqref{eq:utilityUBgeneral} is upper bounded by $t$.

In Step 2 of Algorithm~\ref{alg:mainalgo}, we apply the transformed
Laplace mechanism to every block $v_{(i)} \in \mathbb
R^{\Tdim_{(i)}}$ as described in Section~\ref{apx:generalproofk1}.
However, this requires increasing the sensitivity to $\Delta_{\mathbb
T} \leq \frac{2K}{n}$ due to the increased total number of
measurements. Since the proxy utility loss function $\utilityloss$ is simply
the maximum over the Wasserstein distances over the projected measures
of every subspace $T_{(i)}$ (see
Equation~\eqref{eq:Wassersteinsparse}), we can define the proxy utility loss function $\utilitylossT$ dominating the utility loss $\utilityloss$
(see Definition~\ref{def:utilityloss}) to be the maximum loss
$\utilitylossT(v, v') = \max_{i\in[ K]}
~\utilitylossTi(v_{(i)}, v_{(i)}')$  where $v_{(i)} \in \mathbb
R^{\Tdim_{(i)}}$ is the $i$-th block of the vector $v \in \mathbb
R^\Tdim$ and $\utilitylossTi$ is the corresponding proxy utility loss function as
constructed in Section~\ref{apx:generalproofk1}. We then obtain
the desired result when bounding  the term
$ \mathbb E \utilitylossT(\mathbb T \mu_D, \nudp)$ from
Equation~\eqref{eq:utilityUBgeneral} using
Lemma~\ref{lm:mainappendix}.

\begin{lemma}
\label{lm:mainappendix}
Assume that $\nudp$ is generated as in described in the proof of Theorem~\ref{thm:mainthm}.
We can upper bound $ \utilitylossT(\mathbb T \mu , \nudp)$ from Equation~\eqref{eq:utilityUBgeneral} by:
\begin{equation}
   \mathbb E~\utilitylossT(\mathbb T \mu_D, \nudp) \lesssim  \max_{i \in [K]}   \frac{2 K (\ceil{\log_2(\Tdim_{(i)}) }+1 + \log(K))^2 } {n \epsilon} L_{(i)}.
\end{equation}

 \end{lemma}
\paragraph{Proof of Lemma~\ref{lm:mainappendix}}
Since the sensitivity for every block $i \in [ K]$ is $\frac{2}{n}$ (see Section~\ref{apx:generalproofk1}), we get $\Delta_{\mathbb T} \leq \frac{2K}{n}$. We can then upper bound the expected privacy error when applying the Laplace mechanism:
\begin{equation}
\begin{split}
          \mathbb E ~\utilitylossT(\mathbb T \mu_D, \nudp) &=  \EE \max_{i=[K]} \mathcal{L}_{\mathbb T_{(i)}}(v_{(i)}, v_{(i)} + [\Phi_{(i)} \tilde \eta_i]_{1:\Tdim_{(i)}}) \\
      &= \mathbb E~ \max_{i=[ K]} \sum_{j_{(i)}=1}^{\Tdim_{(i)}} (w_{j_{(i)}+1} - w_{j_{(i)}}) \left\vert \sum_{l=1}^{j_{(i)}} (\Phi_{(i)} \tilde \eta_{(i)})_l\right\vert \\
      &\leq 
     \mathbb E~ \max_{i=[K]}~L_{(i)} ~\max_{j_{i} \in [ \Tdim_{(i)}]} \left\vert \sum_{l=1}^{j_{(i)}} (\Phi_{(i)} \tilde \eta_{(i)})_l\right\vert.
     \label{eq:ubmainupperexp}
\end{split}
\end{equation}
Next, recall that by the construction of the noise $\eta$ in the proof for the case where $K=1$ we have $\tilde \eta_{(i)} \sim \left(\mathrm{Lap}\left(\frac{2 K \|\Phi^{-1}_{{(i)}}\|_1}{n \epsilon}\right)\right)^{\Tdim_{\Phi_{(i)}}}$. In particular, as in Section~\ref{apx:generalproofk1}, we choose $\Phi_{(i)}$ to be the rescaled Haar matrix as described in Lemma~\ref{lm:haar_properties}, and hence
$\tilde \eta_{(i)} \sim \left(\mathrm{Lap}\left(\frac{2 K }{n \epsilon} \right)\right)^{\Tdim_{(i)}}$ with $\Tdim_{\Phi_{(i)}} = 2^{\ceil{\log_2(m_{(i)})}}$. We can now use essentially the following standard argument as in Section 3.3 in \citep{boedihardjo2022private}:

Since for every $i, j_{(i)}$, $ \frac{n \epsilon}{2 K } \tilde \eta_{(i),j}$ has sub-exponential norm $\|\tilde \eta_{(i),j}\|_{\phi_1} \leq 2$ (see Section~2 in \cite{vershynin2018high}), we can apply Bernstein's inequality, which gives together with Lemma~\ref{lm:haar_properties} and the fact that $\|\Phi\|_{\infty} \leq k_{(i)}/2$, for all $i,j_{(i)}$:
\begin{equation}
\begin{split}
    \mathbb P\left(\left| \frac{n \epsilon}{2 K } \sum_{l=1}^{j_{(i)}} (\Phi_{(i)} \eta_{(i)})_l \right| \geq t\right) \leq 2\exp\left(-c\min\left(\frac{t^2}{(k_{(i)}+1)^3}, \frac{t}{k_{(i)}+1}\right) \right) \\
    \leq 2\exp\left(-c\min\left(\frac{t^2}{(k_{\max}+1)^3}, \frac{t}{k_{\max}+1}\right) \right)
  \end{split}
\end{equation}
where $k_{\max} = \max_{i\in[ K]} k_{(i)}$. We can then upper bound the term in Equation~\eqref{eq:ubmainupperexp} when taking the union bound over at most $K \max_{i\in[ K]} \Tdim_{(i)} \leq \exp( \log(K) + k_{\max} + \log(2))$ elements. 
Thus, we obtain the following upper bound for the expectation:
\begin{equation}
\begin{split}
  \mathbb E~ \max_{i \in [ K]} L_{(i)} \max_{j_{(i)} \in [ \Tdim_{(i)}]} \left| \frac{2K}{n \epsilon}\sum_{l=1}^{j_{(i)}} (\Phi_{(i)} \tilde \eta_{(i)})_l \right|  \\
   \lesssim \max_{i \in  [K]\}}   \frac{2 K (\ceil{\log_2(\Tdim_{(i)}) }+1 + \log(K))^2 } {n \epsilon} L_{(i)}.
\end{split}
\end{equation}

\subsection{Proof of Theorem~\ref{cor:rates}}
\label{apx:main}

We now discuss how Theorem~\ref{thm:mainthm} implies Theorem~\ref{cor:rates}. We recall that in this case,  $T_{(i)} =[0,1]^s$ and $\rho_{(i)} = \|. \|_{\infty}$, and thus, for any $k \in \mathbb N_{>0}$, we can simply upper bound the covering numbers by $N(T_{(i)}, \|.\|_{\infty},t) \leq k^s.
$ with $t=1/2k$.
Plugging this upper bound into Equation~\eqref{eq:eq1mainthm} in Theorem~\ref{thm:mainthm} and using that  ${d \choose s} \asymp d^s$, we obtain 
 $   \mathbb E ~\utilityloss(\mu_D, \algoD) \lesssim_s \frac{1}{2k} + \frac{d^s\log(k)^{2}}{n \epsilon} k^{s-1}$, 
 where $\lesssim_s$ is hiding constants depending on $s$. We then obtain the desired result when optimizing over $k$. Finally, we note that the procedure described in the proof of Theorem~\ref{thm:mainthm}  in Appendix~\ref{apx:generalproofkall} and \ref{apx:generalproofk1} agrees exactly with the Algorithm~\ref{alg:mainalgo} from Section~\ref{sec:algo_instantioation} when $T= [0,1]^d$ equipped with the $\ell_{\infty}$ distance function. 

\section{Other types of sparsity: low dimensional data}
\label{apx:measure}

As shown in \citep{Yiyun23,boedihardjo2022private}, the expected utility loss when $\mathcal F$ is the set of all $1$-sparse functions (see Section~\ref{sec:relatedwork}) is of order $\left(\frac{1}{n\epsilon}\right)^{1/d}$. In Section~\ref{sec:mainthm} we showed that we can overcome this curse of dimensionality when restricting $\mathcal F$ to sparse functions. In this section we consider the case where $\mathcal F$ is the set of all $1$-Lipschitz functions and show how this curse of dimensionality can also be overcome when the data lives on a sparse (although unknown) subspace. 
As we show, this is  the case even when we do not have access to any oracle knowledge about the data set, nor the ''dimension" of the subspace, as the algorithm is capable of ``adapting'' to the data set. 

\paragraph{Special case: rates on the hyper cube} Consider the same setting as in Theorem~\eqref{cor:rates}, where the underlying space is the $d$-dimensional hypercube $T =  [0,1]^d$ equipped with the $\ell_{\infty}$-metric. 
Let
$ T_{k,d} = \{ 1/2k, \cdots , (2k-1)/2k\}^d$
 be the centers of a minimal $1/2k$-covering of $T$ of size $N =k^d$. We have:
\begin{theorem}
\label{cor:ratesmeasure}
Let $c\geq 0$ be any constant and let $\mathcal F$ be the set of all $1$-Lipschtiz continuous functions on $[0,1]^d$ with respect to the $\ell_{\infty}$-metric. 
For any $n\epsilon \geq d+1$, there exists an $\epsilon$-DP algorithm $\mathcal A$ such that for any data set $D$,
\begin{equation}
    \label{eq:ratesmeasure}
    \mathbb E ~\utilityloss (\mu_D, \algoD) \lesssim_{s}   \left( \frac{d^3 \log(\epsilon n)}{n \epsilon}\right)^{1/(s+1)} + \frac{d^2 \log(n\epsilon)^2}{n \epsilon^2},
\end{equation}
where $s\in \{0, \cdots, d\}$ is the smallest integer such that for all $k\geq 1$, $D$ is contained in at most $c k^s$ $\ell_{\infty}$-balls of radius   $1/2k$ with centers in $T_{k,d}$.
\end{theorem}
The proof is a consequence of Theorem~\ref{thm:mainthmmeasure} below. 
Note that the definition of $s$ in Theorem~\ref{cor:ratesmeasure} resembles the definition of the Minkowski dimension when the data set $D \subset T_s$ lives on a subspace $T_s$ of Minkowski dimension $s$. The algorithm in Theorem~\ref{cor:ratesmeasure} is ``adaptive" in the sense that it adjusts to the characteristic $s$ of the data set without relying on any prior information or oracle knowledge of $s$. When the worst case scenario occurs and $s = d$, the rate in Equation~\eqref{eq:ratesmeasure} includes an additional term with an exponent of $+1$ compared to Equation~\eqref{eq:ubfull}. This raises the question of whether the cost of adaptivity can be reduced further.

\paragraph{General result}
Generally, for any $t > 0$ and metric space space $\brace{T, \rho}$, fix a
minimal $t$-covering $T_\Tdim$ of size $N(T, \rho, t)$ and let
$\discfunct: T \to  T_\Tdim$ be any measurable discretization map such
that for any point $x \in T$, $\rho(\discfunct(x), x) \leq
t$. Further, let $\vert q_t(D)\vert$ be the size of the support of $q_t(D)$ (which captures the ``sparsity'' of the subspace the data is lying on). We have:
\begin{theorem}
    \label{thm:mainthmmeasure}
In the setting described above, there exists a randomized algorithm
$\gA$ that takes a data set $D \in T^n$ of size $n$  as input on a
metric space $\brace{T, \rho}$ and returns a finitely-supported
probability measure $\f{\gA}{\mu_D}$ on $\brace{T, \rho}$ such that
$\gA$ is $\epsilon$-DP private and has expected utility loss
\eqref{eq:utilityloss} over the set of all $1$-Lipschitz continuous
functions with respect to $\rho$ at most: 
\begin{equation}
\label{eq:ubmeasure}
\mathbb E~\f{\utilityloss}{\mu_D, \f{\gA}{D}} \leq \discdelta +  \frac{64~\mathrm{diam}(T) \log(\Tdim+1)}{n \epsilon}  ~\vert \discfunct(D)\vert 
\end{equation}
\end{theorem}

\subsection{Proof of Theorem~\ref{thm:mainthmmeasure}}
As in Section~\ref{apx:generalproofk1}, let $\mathbb D_{T_\Tdim} : \probset \to
\measure(T_\Tdim)$ be any projection operator and let $\mathbb T\mu$
be any vector on the probability simplex representing the measure $\mathbb D_{T_{\Tdim}} \mu$.
Unlike in Section~\ref{apx:generalproofk1}, we can choose any
random indexing of the elements in $T_\Tdim$ to represent the vector
$\mathbb T \mu$. 

\paragraph{Data sanitization, Step 2 in Algorithm~\ref{alg:mainalgo}:}
We are now going to construct a DP vector $\nudp$ as in Step 2 in
Algorithm~\ref{alg:mainalgo}. A simple
way to construct a sparse private variant of $v$ is: first apply the
standard Laplace mechanism (with $\Phi=I_\Tdim$) to obtain a
differential private copy of $ \tilde{v}_{\text{DP}} = v + \eta$  of
$v$ with $\eta$ as in Lemma~\ref{lm:privacy_phi} and then solve the
convex optimization problem
\begin{equation}
    \nudp = \arg\min_{v'} \|v' - \tilde{v}_{\text{DP}} \|_2 \quad \text{s.t.}~\|v'\|_1 \leq 1
\end{equation}
Standard results for the constrained $\ell_1$-norm ERM solution (see e.g., Theorem 7.13 in \cite{wainwright2019high}) then yield the following upper bound on the $\ell_1$-error $ \|\nudp - v \|_1 \leq 16 \vert\discfunct(D)\vert   \|\eta\|_{\infty}$.

\paragraph{Optimization, Step 3 in Algorithm~\ref{alg:mainalgo}}
For the proxy utility loss we can simply choose $\proxyutility$ form Section~\ref{subsec:tightcertificate}. Note that since $\mathcal F$ is the set of all $1$-Lipschitz queries (and thus $s=d$ in the notation in Section~\ref{subsec:proxy}), we can  solve the minimization problem in Step 3 by solving a linear program (see also~\citep{Yiyun23}).

\paragraph{Upper bound for the utility loss $\utilityloss$:} Recall
from Section~\ref{apx:generalproofk1} that the projection error
$\utilityloss(\mu_D, \mathbb T^\dagger \mathbb T \mu_D) \leq t$ from
Equation~\eqref{eq:utilityUBgeneral} is upper bounded by $t$.  We can
upper bound the (expected) privacy error term  $ \mathbb E \proxyutility(\mathbb
T \mu_D , \nudp)$ in Equation~\eqref{eq:utilityUBgeneral} (where we replace $\utilitylossT$ with $\proxyutility$) by:
\begin{equation}
\begin{split}
   \mathbb E \proxyutility(\mathbb T \mu_D, \nudp) &\leq   \mathbb E \sup_{f\in F; f(0) = 0}  \sum_{z_i \in T_\Tdim} \vert f(z_i) (v_{\mathrm{DP},i} - v_i)\vert \overset{\mathrm{\text{H\"older}}}{\leq}  \mathbb E \sup_{f\in F; f(0) = 0} \|f\|_{L_{\infty}} \|\nudp - v\|_1 \\
     &\leq  \mathrm{diam}(T) ~16 ~\vert \discfunct(D)\vert ~ \mathbb E
    \|\eta\|_{\infty}.
    \label{eq:finalubmainmeasure}
    \end{split}
\end{equation}
We then obtain the desired result when using the upper bound $\mathbb E~\|\eta\|_{\infty} \leq \frac{4}{n \epsilon} \log(\Tdim+1)$ where we used Example 2.19 in \citep{wainwright2019high} in the last line and the fact that  $\|\eta\|_{\psi_1} =2$ for $\eta \sim \mathrm{Lap}(1)$.


\subsection{Proof of Theorem~\ref{cor:ratesmeasure}}
Finally, we discuss how we obtain Theorem~\ref{cor:ratesmeasure} form Theorem~\ref{thm:mainthmmeasure}. Unlike in the proof of  Theorem~\ref{cor:rates} we can no longer simply optimize over $t$ because we do not have access to $\vert  q_{t}(D)\vert$, nor do we assume to have access to the smallest integer $s$ from Theorem~\ref{cor:ratesmeasure} such that for all $t=1/2k$,   $\vert  q_{1/2k}(D)\vert \leq ck^s$.

Instead, we need to ``adaptively'' optimize over all $s' \in \{0, \cdots, d\}$.
For this, in the first step, we want to find for every $s'$  the optimal $t_{s'}$ which minimizes the RHS in Equation~\eqref{eq:ubmeasure} in Theorem~\ref{thm:mainthmmeasure},  assuming that for all $t=1/2k$,   $\vert  q_{1/2k}(D)\vert \leq ck^{s'}$. We choose $t_{s'}= 1/2k_{s'}$ and since $N([0,1]^d, \|. \|_{\infty}, t_{s'}) = k_{s'}^d$, we can upper bound the RHS in Equation~\eqref{eq:ubmeasure} in Theorem~\ref{thm:mainthmmeasure} by:
\begin{equation}
\begin{split}
\label{eq:ubberoundusingks}
    1/2k_{s'} +  \frac{64\log(k_{s'}^d+1)}{n \epsilon}  ~\vert q_{1/2k_{s'}}(D)\vert 
    \leq
    1/2k_{s'} + 1\frac{64 d\log(k_{s'})}{n  \epsilon}  ~c k_{s'}^{s'}.
    \end{split}
\end{equation}
We can now minimize the RHS by choosing  $k_{s'} \asymp \left( \frac{d^2 \log(\epsilon n)}{n \epsilon}\right)^{1/(s'+1)},$  which gives 
\begin{equation}
   1/2k_{s'} +  \frac{64\log(k_{s'}^d+1)}{n \epsilon} \lesssim  \left( \frac{d \log(\epsilon n)}{n \epsilon}\right)^{1/(s'+1)}. 
\end{equation}
We run the algorithm in Theorem~\ref{thm:mainthmmeasure} for every choice of $s' \in \{0, \cdots, d\}$ with $t_{s'}$, resulting in the measures $\mathcal A_{s'}(D)$, and by  Theorem~\ref{thm:mainthmmeasure} we have
\begin{equation}
\label{eq:ub_apx_measure}
    \mathbb E~\utilityloss(\mu_D, \mathcal A_{s'}(D)) \leq 1/2k_{s'} + \frac{64~\log(k_{s'}^d+1)}{n  \epsilon}  ~\vert  q_{1/2k_{s'}}(D)\vert.
\end{equation}
The problem remains which measure $\mathcal A_{s'}(D)$ to return.
The idea is to estimate $\vert  q_{1/2k_{s'}}(D)\vert$ using the estimates $\hat{S}_{s'}$ for the support. More precisely, we release the $d+1$ $\epsilon$-DP estimates for the sizes of the supports:
\begin{equation}
    \hat{S}_{s'} = \vert  q_{1/2k_{s'}}(D)\vert +\frac{1}{ \epsilon} \xi_{s'} ~~\mathrm{with}~~\xi_{s'} \sim \mathrm{Lap}(1)
\end{equation}
where we used that the sensitivity of the support function is trivially $1$. 
We can now return the measure $\mathcal A_{s_{\mathrm{opt}}}(D)$ with 
\begin{equation}
    s_{\mathrm{opt}} = \arg \min_{s'} 1/2k_{s'} +  \frac{64~\log(k_{s'}^d+1)}{n  \epsilon}  ~\hat{S}_{s'}. 
\end{equation}
Note that by the composition theorem for differential privacy \citep{dwork2006calibrating}, the overall algorithm is therefore $2(d+1) \epsilon$-DP, and we obtain an $\epsilon$-DP algorithm by simply replacing $\epsilon$ with $\tilde \epsilon = \epsilon/(2d+1)$.

\paragraph{Upper bound for the expected utility loss:}
To prove the result in Theorem~\ref{cor:ratesmeasure}, we need to upper bound the expected utility loss. We divide the upper bound into two parts, where we let $\mathcal E$ be the event where 
\begin{equation}
    \mathcal E: ~~~~\max_{s' \in \{0, \cdots, d\}} \vert \xi_{s'}\vert \leq 4 \log(n \epsilon), 
\end{equation}
and note that (using $n \epsilon \geq d+1$), $\mathbb P(\mathcal E^c) \leq \frac{1}{(\epsilon n)^2}$. Since the utility loss is at most $1$ (because the transportation cost is at most $\text{diam}(T) =1$), we have that $\mathbb E~ \left[\utilityloss(\mu_D, \mathcal A_{s_{\mathrm{opt}}}(D)) \vert \mathcal E^c \right] \leq 1$. Moreover, we can bound:
\begin{align*}
    \mathbb  E_{\xi,  \eta}~ \utilityloss(\mu_D, \mathcal A_{s_{\mathrm{opt}}}(D)) &\leq \mathbb  E_{\xi,  \eta}~ \left[\utilityloss(\mu_D, \mathcal A_{s_{\mathrm{opt}}}(D)) \vert \mathcal E \right] + P(\mathcal E^c)  \\
    &\leq \mathbb  E_{\xi,  \eta}~ \left[\utilityloss(\mu_D, \mathcal A_{s_{\mathrm{opt}}}(D)) \vert \mathcal E \right] + \frac{1}{(n\epsilon)^2}. 
\end{align*}
Thus, we are only left with bounding the expected utility loss conditioning on  $\mathcal E$. Note that the expectation in Equation~\eqref{eq:ub_apx_measure} is only over $\eta$, and thus:
\begin{equation}\begin{split}
    \mathbb E_{\xi,  \eta}~ &\left[\utilityloss(\mu_D, \mathcal A_{s_{\mathrm{opt}}}(D)) \vert \mathcal E \right] \\
    & \leq  \mathbb E_{\xi}\left[ 1/2k_{s_{\mathrm{opt}}} +  \frac{64~\log(k_{s_{\mathrm{opt}}}^d+1)}{n \tilde \epsilon}  ~\vert q_{1/2k_{s_{\mathrm{opt}}}}(D)\vert \big \vert \mathcal E\right] \\
    & \leq  1/2k_{s} +  \frac{64~\log(k_{s}^d+1)}{n \tilde \epsilon}  ~\left(\vert q_{1/2k_{s}}(D)\vert + 4 \frac{\log(\epsilon n)}{\tilde \epsilon}\right)  \\
 &\lesssim  \left( \frac{d^2 \log(\epsilon n)}{n \epsilon}\right)^{1/(s+1)} +  \frac{d^3 \log^2(n\epsilon)}{n \epsilon^2},
 \end{split}
\end{equation}
where we used in the last line the assumption that  for all $k$, $\vert q_{1/2k_{s'}}(D)\vert \leq ck^s$ and Equation~\eqref{eq:ubberoundusingks} and recall that $k_{s} \asymp \left( \frac{d^2 \log(\epsilon n)}{n \epsilon}\right)^{1/(s+1)}$.

 \end{document}